\documentclass[journal,onecolumn,10pt]{IEEEtran}
\usepackage{indentfirst}
\usepackage{color}

\newtheorem{Theorem}{Theorem}

\newtheorem{Definition}{Definition}

\newtheorem{Remark}{Remark}
\newtheorem{Lemma}{Lemma}

\newtheorem{Example}{Example}

\usepackage{times,amsmath,epsfig}
\allowdisplaybreaks[4]

\usepackage{verbatim}
\usepackage{caption}
\usepackage{cite}

\usepackage{algorithm}
\usepackage{algpseudocode}
\usepackage{amsmath}
\usepackage{multirow}
\usepackage{graphicx,color}
\usepackage{epstopdf}
\usepackage{graphicx} 
\usepackage{subfigure}
\usepackage{times,epsfig}
\usepackage{float}
\usepackage{bm}
\usepackage{amssymb}
\renewcommand{\algorithmicrequire}

\usepackage{pgfplots}

\usepackage[justification=centering]{caption}

\begin{document}

\title{Adaptive Gradient Coding}
\author{Hankun Cao, Qifa Yan, \IEEEmembership{Member, IEEE,} Xiaohu Tang, \IEEEmembership{Senior Member, IEEE}, Guojun Han, \IEEEmembership{Senior Member, IEEE}
\thanks {H. Cao, Q. Yan and X. Tang are with Information Security and National Computing Grid Laboratory, Southwest Jiaotong University, Chengdu, China. Q. Yan has previously been with the Electrical and Computer Engineering Department, University of Illinois at Chicago, Chicago, IL 60607 USA. (email: hkcao@my.swjtu.edu.cn, qifayan@swjtu.edu.cn, xhutang@swjtu.edu.cn).}

\thanks {G. Han is with the School of Information Engineering, Guangdong University of Technology, Guangzhou, China (email:  gjhan@gdut.edu.cn).}
}

\maketitle

\begin{abstract}
This paper focuses on mitigating the impact of stragglers in distributed learning system. Unlike the existing results designated for a fixed number of stragglers, we develop a new scheme called \emph{Adaptive Gradient Coding (AGC)} with flexible communication cost for varying number of stragglers. Our scheme gives an optimal tradeoff between computation load, straggler tolerance and communication cost by allowing workers to send multiple signals sequentially to the master. In particular, it can minimize the communication cost according to the unknown real-time number of stragglers in  practical environments. In addition, we present a \emph{Group AGC (G-AGC)} by combining the group idea with AGC to resist more stragglers in some situations. The numerical and simulation results demonstrate that our adaptive schemes can achieve the smallest average running time. 
\end{abstract}

\section{Introduction}
Nowadays, large-scale machine learning and data analysis tasks are increasingly run on the modern distributed computing platforms. Typically such platform is comprised of huge number of computing nodes (also called workers), where the performance is significantly affected by the behavior of some unreliable workers\cite{DeanB13}. For example, some workers can introduce a delay of up to $5\times$ the typical performance, which are  called  stragglers  in the literature.
As a result, the non-straggler workers should try to communicate with the master node in an efficient way to proceed the computation procedure. That is, the system designer has to consider both straggler tolerance and communication cost.

To handle stragglers, coding theory has been effectively introduced into distributed computing system with techniques from  distributed storage \cite{DimakisGWWR10} and coded caching \cite{Maddah-AliN15}.
In \cite{LeeLPPR18}, Lee \emph{et al.} first proposed an erasure-correcting code to mitigate the effects of stragglers by regarding them as erasures in machine learning tasks, such as linear regression and matrix multiplication. Subsequently, many codes were designed to facilitate various practical situations. For instance, in oder to reduce the computation load, ``ShortDot"   \cite{DuttaCG19} and ``Sparse LT" \cite{WangLS18_sparse} codes were respectively presented in matrix multiplications; Polynomial codes were constructed to increase the tolerance ability  of stragglers in high-dimension matrix multiplication problems \cite{YuMA17,YuMA18,Dutta18}; Distributed computing schemes based on placement delivery array  \cite{YanPDA} were developed for map-reduce systems \cite{Yan20,Kumar19}; 
The utilization of non-persistent stragglers, i.e., workers that can finish part of the assigned works, was investigated in \cite{OzfaturaGU19,KianiFD18,DasTR18}.

Among all the computation tasks, gradient based optimization algorithms are a class of most important ones used in distributed machine learning systems. In \cite{TandonLDK17}, Tandon \emph{et al.} first formalized a coding framework called \emph{gradient coding} to combat with persistent stragglers for gradient calculation in distributed systems. By the technique of adding redundant computing tasks at the distributed workers, the master only needs to wait for a subset of results and then can ignore the stragglers. Consequently, it achieves a flexible tradeoff between computation load and straggler tolerance. Later on, several other codes were  constructed to achieve the tradeoff \cite{HalbawiRSH18,RavivTDT18}. 
Unlike all these works applying the same code to all coordinates of gradient vectors, Ye \emph{et al.} proposed to encode across different coordinates of gradients to reduce the size of transmitted vectors, which results in a flexible and optimal tradeoff between computation load, straggler tolerance, and communication cost \cite{YeA18}. 
 The resultant optimal communication cost is crucial in practical environments, since the gradient normally consists of millions of real-valued coordinates, and sometimes the transmission of these high-dimensional vectors can amortize the savings of the distributed computation \cite{LiASY14}. In \cite{Kadhe20}, the authors used group idea to obtain the same tradeoff with lower complexity and better numerical stability compared to the scheme in \cite{YeA18}. Very recently, based on history information of the straggler behavior, an optimization framework of grouping was presented in \cite{Buyukates20} to minimize the completion time. However, it introduces many overheads of feedbacks and thus is not appropriate when stragglers change rapidly among iterations.
 
 Besides, there are other coding schemes to cope with non-persistent stragglers and approximate gradient. 
 When non-persistent stragglers are considered, computation scheduling schemes were proposed in \cite{Amiri19,Ozfatura20}, where the computation load of each worker can be reduced at the expense of  increased communication cost, i.e., transmitting multiple full-size gradient vectors. Nevertheless, it might be unacceptable in practice when the gradient vector is of high-dimensions and the communication bandwidth is limited.
In \cite{RavivTDT18,WangLS18,HoriiYKM19,KadheKR19}, instead of recovering the exact gradient, the works  considered recovering an approximate gradient in each iteration and showed that one can still obtain a good approximation of the original solution by using coding techniques.  A speedup was obtained at the cost of an increased generalization error. 

 In addition to coding methods, some efforts on asynchronous approaches were made recently to tackle the straggler issues in learning system \cite{LiASY14,HoCCLKGGGX13,MitliagkasZHR16,DuttaJGDN18}. While it is a valid way to deal with straggler workers, it gives up many other desirable properties such as convergence rate, amenability to analysis, ease reproducibility and so on \cite{TandonLDK17}.

 In this paper, we concentrate on the same scenario  as those in \cite{TandonLDK17,YeA18,Kadhe20}.  Precisely,  we aim at designing a new type of gradient coding with exact gradient recovery in each iteration without any history information about the behavior of persistent stragglers. Note that all the aforementioned codes for gradient descent are designed for a fixed number of stragglers.
Whereas in  most practical environments,  the real-time number of stragglers is unknown in advance, thus generally what one can do is to estimate the maximal number of stragglers in the worst case. Moreover, the real-time number of stragglers may vary over iterations. Then, the codes designed for the worst case may deteriorate the performance when there are less stragglers.
Therefore, we investigate a more realistic gradient coding setup, where the master can decode with the least communication cost for \emph{any} tolerable number of stragglers
  less than a threshold. Particularly, a new coding scheme called \emph{Adaptive Gradient Coding (AGC)} is proposed to meet the requirement, i.e., the proposed code can achieve the minimal communication cost depending on the real-time number of stragglers even without its knowledge in advance, which can then obtain a better performance compared to the schemes in \cite{TandonLDK17,YeA18}. 
 We support our theoretical results by implementing AGC on Amazon EC2 clusters using Python with mpi4py package. The Cifar-10 dataset is trained with ResNet-18, and the experimental results show that the proposed scheme can be used to tackle varying number of stragglers with smaller average running time compared to \cite{TandonLDK17,YeA18}, while maintaining the same test accuracy. 


\emph{Notations:} We use  $\mathbb{N}$, $\mathbb{N}^+$ and $\mathbb{R}$ to denote the set of non-negative integers, positive integers and real numbers respectively. For any $a,b\in\mathbb{N}$ such that $a<b$, we use $[a:b]$  to denote the set $\{a,a+1,\ldots,b\}$, and $[a:b)$ to denote the set $\{a,a+1,\ldots,b-1\}$. The notation $\bm 1_{a\times b}$ and $\bm 0_{a\times b}$ are used to denote a matrix of size $a\times b$ with all entries being $1$ and $0$ respectively. For convenience, we start all the coordinates of vectors and matrices with $0$. For a matrix $M\in\mathbb{R}^{a\times b}$ and index sets $\mathcal{H}\subseteq[0:a),\mathcal{P}\subseteq[0:b)$, we denote the sub-matrix $M_{\mathcal{H},:}$  formed by the rows in $\mathcal{H}$, $M_{:,\mathcal{P}}$  formed by columns in $\mathcal{P}$, and $M_{\mathcal{H},\mathcal{P}}$  formed by rows in $\mathcal{H}$ and columns in $\mathcal{P}$, where the orders of rows/columns are  consistent with those in $M$.

\section{Problem Formulation and Related Works}
\subsection{Problem Setup}\label{subsec:problem}
Consider a system consisting of a master and $n$ distributed workers, which aims to train a model from a set of $N$ data points $\mathcal{D}=\{(x_i,y_i)\}_{i=0}^{N-1}$, where $x_i\in\mathbb{R}^{w}$ is the $i$-th observation, and $y_i\in\mathbb{R}$ is the $i$-th response. 
 The training is performed by solving an optimization problem
\begin{eqnarray*}
	\beta^{\ast} = \arg\min_{\beta} \sum_{i=0}^{N-1}\ell(x_i,y_i;\beta),
\end{eqnarray*}
where $\ell(\cdot)$ is a given per-sample loss function, and $\beta\in\mathbb{R}^w$ is the parameter to be optimized. In general, this optimization problem can be solved with a gradient-based algorithm. More precisely, given an initial estimation $\beta^{(0)}$, the gradient-based algorithm produces a new estimation $\beta^{(t)}$ in each iteration $t=1,2,\ldots$, i.e.,
\begin{eqnarray*}
	\beta^{(t+1)} = \mathcal{R}(\beta^{(t)},g^{(t)}),
\end{eqnarray*}
where $\mathcal{R}$ is the update function and
\begin{eqnarray}
  g^{(t)}\triangleq\sum_{i=0}^{N-1}\nabla \ell (x_i,y_i;\beta^{(t)})\in\mathbb{R}^w\label{eqn:gradient:gt}
\end{eqnarray}
is the \textit{gradient} of the loss function at the $t$-th estimate $\beta^{(t)}$.

For large $N$, the bottleneck of the system is the calculation of the gradient in \eqref{eqn:gradient:gt}, which is usually run by dispersing the tasks across $n$ workers. Suppose that each worker can store up to $\mu N$ data points for some $\mu\in[\frac{1}{n},1]$ \footnote{Notice that, for $\mu<\frac{1}{n}$, the total memory of all the workers is insufficient to store all the data points at least once; while for $\mu>1$, all the data points can be stored at any given worker.}. For simplicity, we  always assume that $\mu N\in\mathbb{N}^{+}$  in this paper.
We call the above system as a $(n,\mu,w)$ distributed gradient computing system, or $(n,\mu,w)$ system for short.

In principle, a distributed gradient computing scheme is carried out as follows.
Firstly, the data set $\mathcal{D}$ is partitioned into $k$ equal-size subsets $\mathcal{D}_0,\ldots,\mathcal{D}_{k-1}$ for some $k\in\mathbb{N}^+$. 
Each worker $j\in[0:n)$ stores $d$ subsets $\{\mathcal{D}_{j_0},\ldots,\mathcal{D}_{j_{d-1}}\}$ for some $d\in[1:k]$, where $j_0,j_1,\ldots,j_{d-1}$ are $d$ distinct indices in $[0:k)$. The memory restriction at the workers imposes the constraint on the choice of $k$ and $d$:
\begin{eqnarray} \label{storage:constraint}
  \frac{d}{k}\leq \mu.
\end{eqnarray}
For any $i\in[0:k)$, let $g_i\in\mathbb{R}^{w}$ be the partial gradient\footnote{Since we focus on one iteration of the updates, we will omit the superscript ``$(t)$" for simplicity in the following in this paper.} over data subset $\mathcal{D}_i$, i.e.,
\begin{eqnarray}
g_i \triangleq \sum_{(x,y)\in\mathcal{D}_i}\nabla \ell(x,y;\beta).\label{eqn:gr}
\end{eqnarray}

Clearly, according to \eqref{eqn:gradient:gt} and the fact that $\mathcal{D}_0,\ldots,\mathcal{D}_{k-1}$ form a partition of $\mathcal{D}$, the gradient  $g$ can be expressed as
 \begin{eqnarray} \label{eqn:gradient:g}
 g=\sum_{i=0}^{k-1} g_i.
 \end{eqnarray}

Next, each worker $j\in[0:n)$ computes their own $d$  partial gradients $g_{j_0},g_{j_1} ,\ldots,g_{j_{d-1}}$ respectively.
Suppose that, in each iteration, within given time threshold only $n-s$ workers are able to finish their computation tasks, which are called \emph{active workers, where $s$ is the number of stragglers}. 

Once accomplishing the computation of the $d$ partial gradients, each active worker $j$ starts to calculate at most $q_{s_{\max}}$ coded signals
\begin{eqnarray*}
f_{j,l}(g_{j_0},g_{j_1},\ldots,g_{j_{d-1}}),\quad l\in[0:{q_{s_{\max}}}),
\end{eqnarray*}
and sends them to the master sequentially, where $s_{\max}$ is the largest number of stragglers that a scheme can tolerate, and $f_{j,l}:\mathbb{R}^{dw}\mapsto \mathbb{R}$ are some linear functions. In the presence of $s$ stragglers, let $\mathcal{F}\subseteq[0:n)$ be the set of active workers where $|\mathcal{F}|=n-s$, the master must decode the full gradient  with the first $q_s$ signals from each of the active workers, i.e.,
 \begin{IEEEeqnarray}{c}
\{f_{j,l}:j\in\mathcal{F},l\in[0:q_s)\}.\label{eqn:signal}
\end{IEEEeqnarray}
We refer to $q_s$ as \emph{communication length} and $q_s\over w$ as \emph{communication cost} in the presence of $s$ stragglers. Note that the real-time number of stragglers $s$ and  active set $\mathcal{F}$ are not known in advance, and thus  unavailabe at each individual worker.

Let $t_i$ be the number of workers storing the data subset $\mathcal{D}_i$. To resist \emph{any }subset of $s$  stragglers, $\mathcal{D}_i$ has to be stored by at least $s+1$ workers, i.e., $t_i\geq s+1$. If $s\geq \lfloor n\mu\rfloor$,  then
	\begin{eqnarray*}
		nd=\sum_{i=0}^{k-1}t_i\geq(s+1)k\geq (\lfloor n\mu\rfloor+1)k>n\mu k,
	\end{eqnarray*}
	which contradicts \eqref{storage:constraint}. 
 Therefore, 
we always assume  $s\leq \lfloor n\mu\rfloor-1$.
Then,  $s_{\max}\leq \lfloor n\mu\rfloor-1$ for any $(n,\mu,w)$ distributed gradient scheme.

The communication cost of the system is characterized by the communication vector defined as follows.

\begin{Definition}[Communication vector]\label{def:achievable}
For a given $(n,\mu,w)$ system, for any $s_{\max}\in[0:\lfloor n\mu\rfloor)$, the communication vector $\bm c = {1\over w}\cdot(q_0,q_1,\ldots,q_{s_{\max}})$ is said to be achievable for $s_{\max}$ stragglers, if there exists $k,d\in\mathbb{N}^+$ satisfying \eqref{storage:constraint} and a corresponding storage design as well as a set of linear encoding functions $\{f_{j,l}:j\in[0:n),l\in[0:q_{s_{\max}})\}$   such that for each $s\in[0:s_{\max}]$, for any active set $\mathcal{F}\subseteq[0:n)$ of size $n-s$, the master can decode the full gradient \eqref{eqn:gradient:g} from the coded signals in \eqref{eqn:signal}.
\end{Definition}

\begin{Definition}[Optimal communication vector] For any two communication vectors $\bm c=\frac{1}{w}\big(q_0,q_1,\ldots,q_{s_{\max}}\big)$ and $\bm c'=\frac{1}{w}\big(q_0',q_1',\ldots,q_{s_{\max}'}'\big)$, we say that $\bm c$ is superior to $\bm c'$ if $s_{\max}\geq s_{\max}'$ and $q_s\leq q_s'$ for all $s\in[0:s_{\max}']$. An achievable communication vector $\bm c$ is said to be optimal if $\bm c$ is superior to any achievable communication vector.
\end{Definition}

The main objective of this paper is to characterize all the achievable and optimal communication vectors 
for any $(n,\mu,w)$ system.

\subsection{Related Works}
The most relevant works to ours are the distributed gradient computing schemes in \cite{TandonLDK17} and \cite{YeA18}. In particular, it was proved in \cite{TandonLDK17} that the communication vector
\begin{IEEEeqnarray}{c}\label{eqn:GC:c}
\bm{c}_{s_{\max}}^{\textnormal{GC}}=(\underbrace{1,1,\ldots,1}_{s_{\max}+1})\label{eqn:c}
\end{IEEEeqnarray}
is achievable for at most $s_{\max}$ stragglers whenever
\begin{eqnarray}\label{eqn:CGC:c}
\frac{s_{\max}+1}{n}\leq \mu.\label{eqn:bound:1}
\end{eqnarray}
Moreover, the conclusion was extended in \cite{YeA18} to that, for any fixed $q$ with $q|w$, the communication vector
\begin{IEEEeqnarray}{c}
\bm c_{s_{\max},q}^{\textnormal{CGC}}=\Big(\underbrace{\frac{q}{w},\ldots,\frac{q}{w}}_{s_{\max}+1}\Big)\label{eqn:c:q}
\end{IEEEeqnarray}
is achievable for  at most $s_{\max}$ stragglers whenever
\begin{IEEEeqnarray}{c}
\frac{s_{\max}+{w \over q}}{n}\leq \mu. \label{eqn:bound:2}
\end{IEEEeqnarray}

In addition, the authors in \cite{Kadhe20} proposed a low complexity group scheme  achieving the same communication vector in \eqref{eqn:c:q} and the bound in \eqref{eqn:bound:2}. It was showed that the scheme has a better numerical stability and smaller complexity compared to the one in \cite{YeA18}, with a constraint  $d\,|\,n$.

In this paper, we refer to the codes in \cite{TandonLDK17} and \cite{YeA18} as \emph{Gradient Coding (GC)} and \emph{Communication-efficient Gradient Coding (CGC)} respectively. 
It can be seen in \eqref{eqn:bound:2} with $q=w$, CGC  degrades to GC as in \eqref{eqn:bound:1}, i.e., GC is a special case of CGC. Therefore, we 
will take CGC with various $s_{\max}$ as baseline in this paper.
Notice that CGC is designed to tackle up to $s_{\max}$ stragglers with fixed communication cost. Consequently, it maintains the same communication cost even when the real-time number of  stragglers is smaller than $s_{\max}$ (see \eqref{eqn:c:q}). In this paper, the proposed AGC scheme has the property that the communication cost can be reduced according to any real-time number of stragglers $s\le s_{\max}$, and thus achieves the optimal communication vector.  We will first discuss the AGC scheme, and combine it with the group method in Section \ref{sec:Group}.

\section{Main Result}

Before presenting the main result, we begin with an example to illustrate the basic idea of AGC scheme.
\begin{Example} \label{eg:motivation}
Consider a $(n,\mu,w)=\big(3,\frac{2}{3},2\big)$ system with $k=3$. We illustrate the codes of CGC and AGC with $s_{\max}\in \{0,1\}$ in Fig. \ref{fig:a}, \ref{fig:b} and \ref{fig:c} respectively. We see that both CGC with $s_{\max}=0$ and AGC apply coding across the coordinates of gradient vector, while CGC with $s_{\max}=1$ does not, e.g., ${5\over 2}g_1(0)+g_0(1)+{1\over 2}g_1(1)$ for CGC($s_{\max}=0$)/AGC and $g_0(0)+2g_1(0)$ for CGC($s_{\max}=1$).
In CGC scheme, if $s_{\max}=1$, each worker transmits a signal of length $2$ to the master, the master decodes the full gradient as long as it receives the signals from any two workers, i.e., it achieves the communication vector $\bm c= (1,1)$ for at most $1$ straggler.
If $s_{\max}=0$, each worker transmits a signal of length $1$ to the master, but the master needs to collect the signals from all the three workers, so the communication cost is ${1\over 2}$.
Thus, CGC with $s_{\max}=1$ can resist one straggler but has higher communication cost $1$, whereas CGC with $s_{\max}=0$ has lower communication cost ${1\over 2}$ but is fragile to any worker failure.
In our AGC scheme, the workers send the signals in at most two rounds depending on the exact real-time number of stragglers: if there is no straggler ($s=0$), after receiving the signals in the first round from the three workers, the master broadcast a stop message and then decode the full gradient; if there is one straggler ($s=1$), the active workers will need to transmit the signals in two rounds to the master.
Therefore, communication vector of AGC scheme is $\bm c = ({1\over 2},1)$, which achieves the corresponding optimal communication cost of CGC  when the real-time number of stragglers equals $s_{\max}$.

	\begin{figure*}[ht]
		\begin{center}
		\subfigure[CGC with $s_{\max}=1$: it can tolerate at most $1$ straggler.] { \label{fig:a}
			\includegraphics[width=0.4\textwidth]{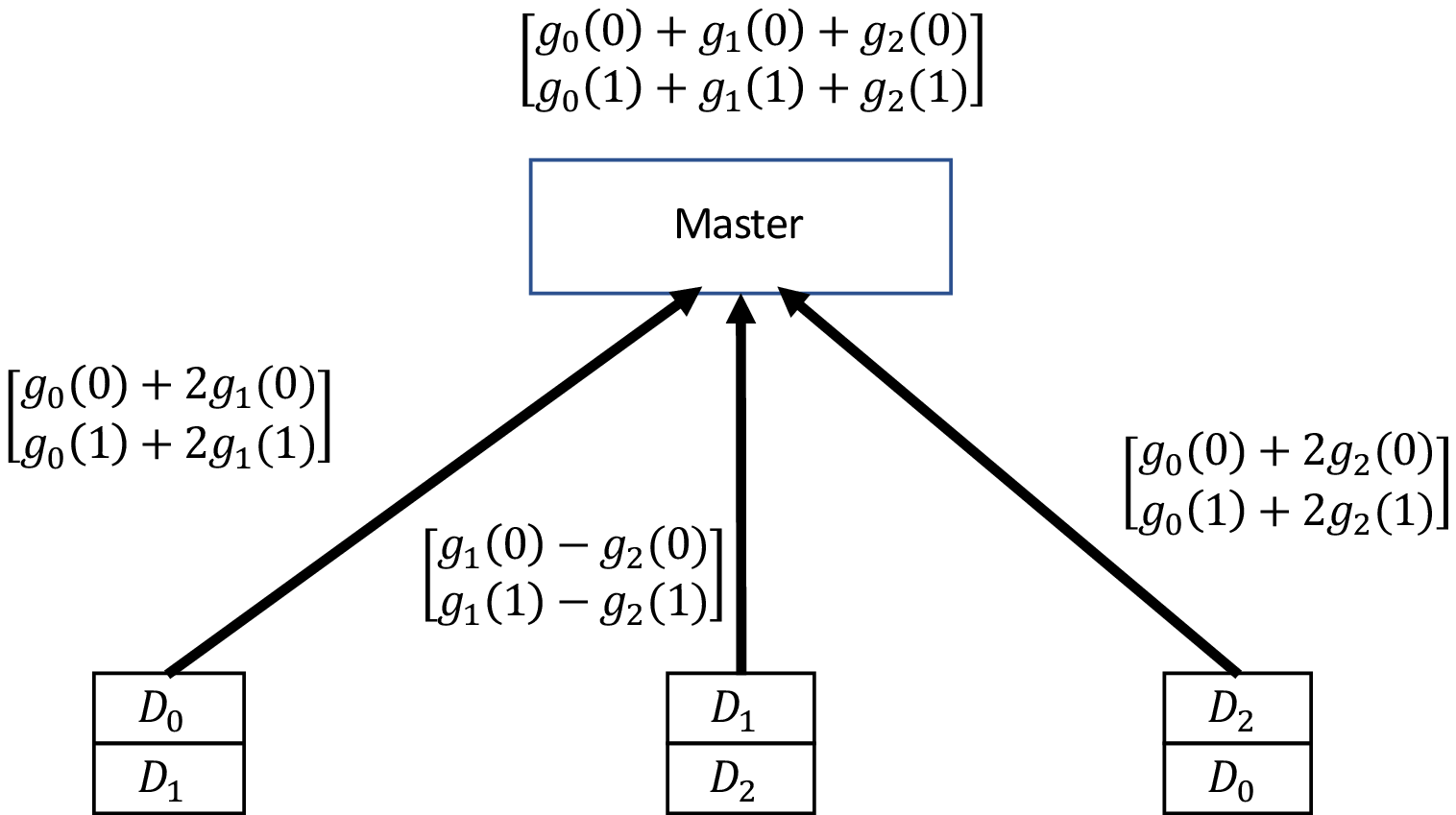}
			}
		\subfigure[CGC with $s_{\max}=0$: it can minimize the communication cost but is fragile to any worker failure.] { \label{fig:b}
			\includegraphics[width=0.5\textwidth]{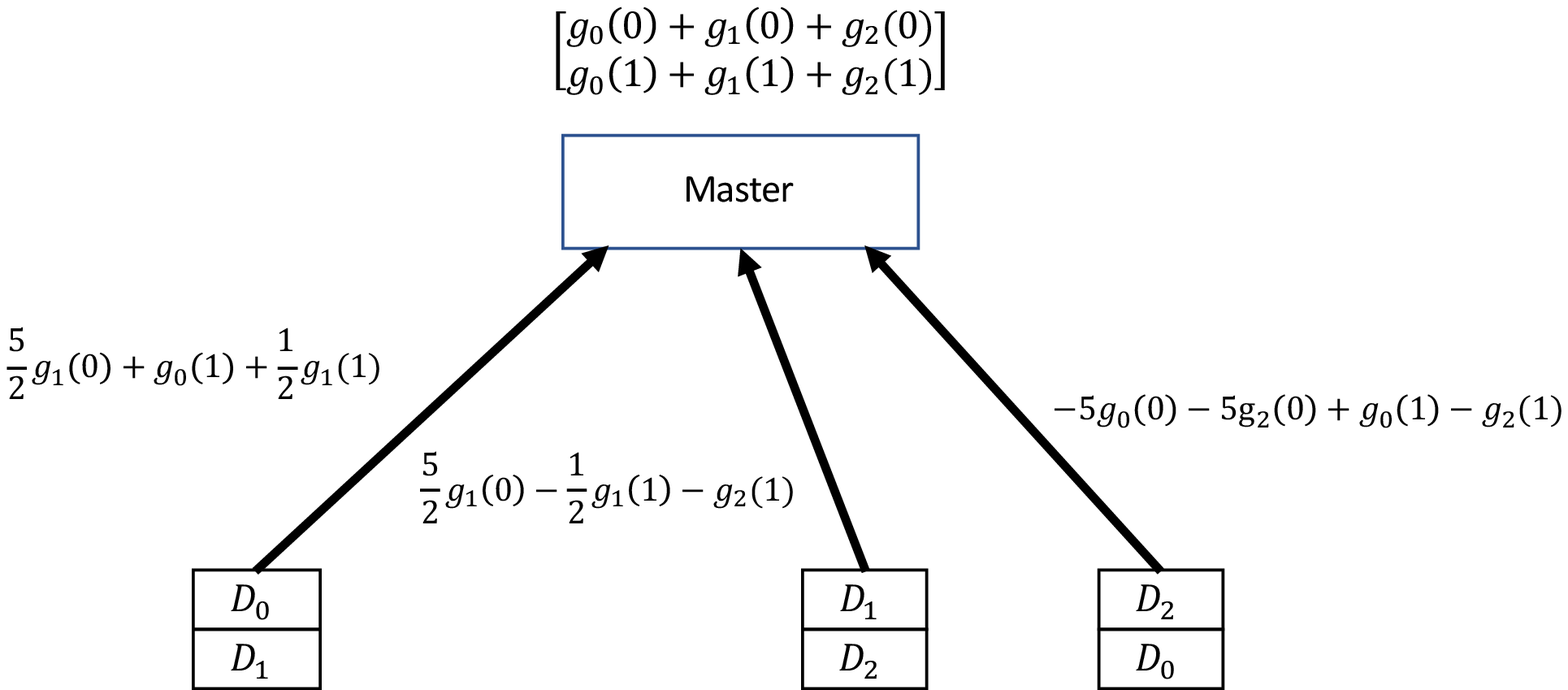}
			}	
		\subfigure[AGC: it can achieve the minimal communication cost according to the real-time number of stragglers with each worker transmitting at most two coded symbols.] {\label{fig:c}
			\includegraphics[width=0.8\textwidth]{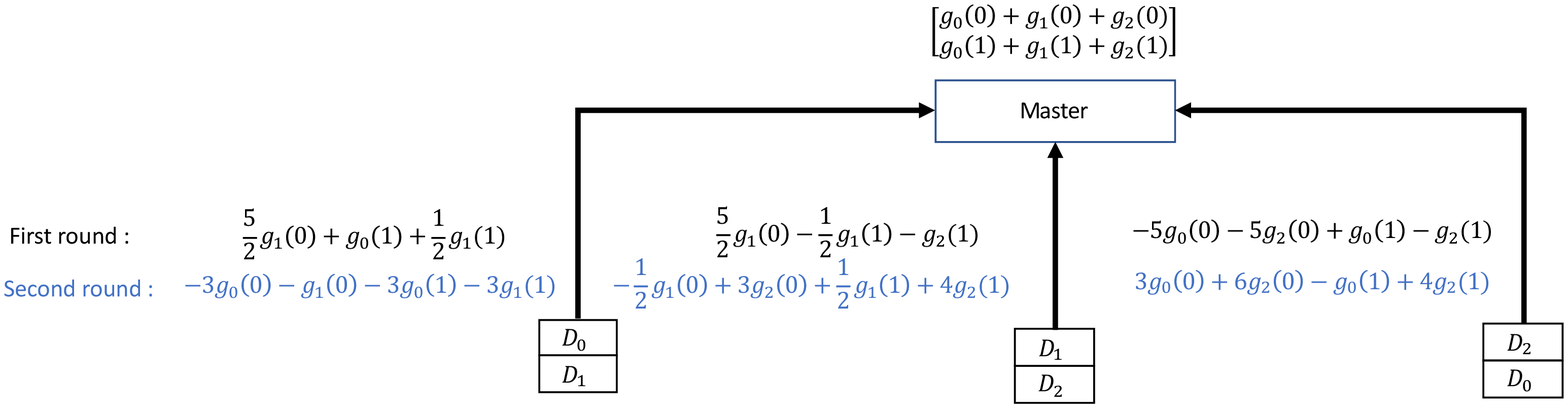}
			}
		\caption{The basic idea of CGC and AGC schemes } \label{fig:motivate_example}
		\end{center}
	\end{figure*}

\end{Example}

\begin{Theorem} \label{theorem:basic}
For a given $(n,\mu,w)$ distributed gradient computing system, where $n,w\in\mathbb{N}^+, \mu\in[\frac{1}{n},1]$, for any $s_{\max}\in[0:\lfloor n\mu\rfloor)$, a communication vector $\bm c = {1\over w}\cdot(q_0,q_1,\ldots,q_{s_{\max}})$ is achievable if and only if
\begin{eqnarray} \label{eqn:thm}
\lfloor n\mu\rfloor\geq s + \Big\lceil\frac{w}{q_s}\Big\rceil,\quad\forall \, s\in\{0,1,\ldots,s_{\max}\}.
\end{eqnarray}
Moreover, the optimal communication vector is given by
\begin{IEEEeqnarray}{c}
\bm c^*=\frac{1}{w}\bigg(\bigg\lceil \frac{w}{\lfloor n\mu\rfloor}\bigg\rceil, \bigg\lceil \frac{w}{\lfloor n\mu\rfloor-1}\bigg\rceil,\ldots,w\bigg).\label{eqn:c:opt}
\end{IEEEeqnarray}
\end{Theorem}

We will show the converse in Section \ref{sec:converse}, and prove the achievability by presenting the general AGC scheme in Section \ref{sec:scheme}.

\begin{Remark}
Unlike $\bm c_{s_{\max},q}^{\textnormal{CGC}}$ in \eqref{eqn:c:q}, $\bm c^*$ in \eqref{eqn:c:opt} has different values in its different coordinates.
It can be easily verified that the optimal vector $\bm c^*$  is superior to $\bm c_{s_{\max},q}^{\textnormal{CGC}}$  for any $s_{\max}\in[0:\lfloor n\mu\rfloor)$ when the signal length $q$ dividing $w$.
This is to say, AGC scheme outperforms CGC schemes in the sense that it adaptively reduces the communication cost according to the real-time number of stragglers.
\end{Remark}

Further, to clearly illustrate the advantages of adaptive performance, we present another example (Example 2).  Compared with the one in Example \ref{eg:motivation}, it shares the same idea but with little bigger parameters. Due to the limited space, in Example \ref{eg:comp} we omit the details of the code and only compare the communication cost.
\begin{Example}\label{eg:comp} Consider a $(n,\mu,w)=(5,\frac{4}{5},12)$ system. According to \eqref{eqn:c:q} and \eqref{eqn:c:opt}, we list the results in Table \ref{table:eg:comp} to compare the communication costs of CGC and AGC codes for different number of stragglers $s=0,1,2,3$, where the cross symbol ``$\times$" indicates that the code fails in the corresponding case.
\begin{table*}[!http]
	\center
	\caption{Communication costs of CGC and AGC for a $(n,\mu,w)=(5,\frac{4}{5},12)$ system.}\label{table:eg:comp}
	\begin{small}
		\begin{tabular}{|c|c|c|c|c|c|c|}\hline
			Number of Stragglers $s$ & CGC  ($q=12$) & CGC ($q=6$) & 	CGC ($q=4$)	&	CGC ($q=3$)	&	AGC\\\hline
0&	1	&	1/2	&	1/3	&	1/4	&	1/4	\\\hline
1&	1	&	1/2	&	1/3	&	$\times$	&	1/3	\\\hline
2&	1	&	1/2	&	$\times$	&	$\times$	&	1/2	\\\hline
3&	1	&	$\times$	&	$\times$ 	&	$\times$	&	1	\\\hline
		\end{tabular}
		\end{small}
	\end{table*}
From Table \ref{table:eg:comp}, AGC can achieve the best communication cost regardless of the real-time number of stragglers.  For any given CGC scheme, it either has higher communication cost, 
 or fails to work.

\end{Example}

\section{Converse and Optimality}\label{sec:converse}

In this section, we prove the converse and the optimality of $\bm c^*$. The converse follows similar lines as in \cite{YeA18}. However, due to different setup and result,  we present it here for completeness.
\subsection{Converse}\label{subsec:converse}
Suppose that the communication vector $\bm c={1\over w}\cdot(q_0,q_1,\ldots,q_{s_{\max}})$ is achievable for at most $s_{\max}$ stragglers where $s_{\max}\in[0:\lfloor n\mu\rfloor)$. Then by Definition \ref{def:achievable}, there exists $k,d\in\mathbb{N}^+$ satisfying \eqref{storage:constraint} such that the data set $\mathcal{D}$ is partitioned into equal-size subsets $\mathcal{D}_0,\mathcal{D}_1,\ldots,\mathcal{D}_{k-1}$, and an assignment such that each worker stores $d$ data subsets, as well as a set of functions $\{f_{j,l}:j\in[0:n),l\in[0:q_{s_{\max}})\}$ satisfying the required decoding conditions, i.e., the master can always decode the full gradient \eqref{eqn:gradient:g} from the signals in \eqref{eqn:signal} for any $\mathcal{F} \in[0:n)$ with $|\mathcal{F}|=n-s$ for any $s\in[0:s_{\max}]$.

Given a data subset $\mathcal{D}_i$, $i\in[0:k)$,  W.L.O.G. assume that it is stored at workers $0,1,\ldots,t_i-1$. Consider the case that the first $s$ workers $0,1,\ldots,s-1$ are stragglers. By definition, the full gradient $g=\sum_{i=0}^{k-1}g_{i}$ can be decoded from the signals
\begin{eqnarray*}
\{f_{j,l}:j\in[s:n),l\in[0:q_s)\}.
\end{eqnarray*}
Since $g_i$ can only be computed from the data subset $\mathcal{D}_i$, which is stored only at the workers $0,1,\ldots, t_i-1$, the partial gradient $g_i$ can then be calculated from
	\begin{eqnarray*}
		\{f_{j,l}:j\in[s:t_i),l\in[0:q_s)\}.
	\end{eqnarray*}
Clearly, to recover the $w$-dimensional vector $g_i$, the number of equations in   $\{f_{j,l}:j\in[s:t_i),l\in[0:q_s)\}$  is at least $w$, i.e.,
\begin{eqnarray*}
q_s(t_i-s)\geq w.
\end{eqnarray*}
Therefore, the integer  $t_i$ satisfies
\begin{eqnarray} \label{Eqn_sa}
t_i\ge s+\Big\lceil\frac{w}{q_s}\Big\rceil,\quad \forall\, i\in[0:k).
\end{eqnarray}
Summing \eqref{Eqn_sa} over $i\in[0:k)$, we get
\begin{eqnarray*}
k\left(s+\Big\lceil\frac{w}{q_s}\Big\rceil\right)\leq \sum_{i=0}^{k-1}t_i= nd.
\end{eqnarray*}
Thus,  by the fact that $s+\big\lceil\frac{w}{q_s}\big\rceil$ is an integer, we have
\begin{eqnarray} \label{eqn:d:size}
s+\Big\lceil\frac{w}{q_s}\Big\rceil\leq\Big\lfloor n\frac{d}{k} \Big\rfloor\leq \lfloor n\mu\rfloor,\quad \forall\, s\in[0:s_{\max}].
\end{eqnarray}
where the last inequality holds due to the constraint \eqref{storage:constraint}.

\subsection{Optimality of The Communication Vector $\bm c^*$}\label{subsec:optimality}
In Section \ref{sec:scheme}, we will show the achievability of $\bm c^*$ in \eqref{eqn:c:opt}. Here, we prove that $\bm c^*$ is superior to any achievable communication vector. Consequently, the optimal communication vector is unique and given by $\bm c^*$.

  Let $\bm c=\frac{1}{w}\cdot(q_0,q_1,\ldots,q_{s_{\max}})$ be any achievable communication vector for $s_{\max}\in[0:\lfloor n\mu\rfloor)$ stragglers. Recall that the largest number of stragglers the system can resist with the  communication vector $\bm c^*$  is $s_{\max}^*=\lfloor n\mu\rfloor -1\geq s_{\max}$. Moreover, by the converse proof, for any $s\in[0:s_{\max}]$,
\begin{IEEEeqnarray}{c}
\lfloor n\mu\rfloor \geq s+\Big\lceil\frac{w}{q_s}\Big\rceil\geq s+\frac{w}{q_s}.
\end{IEEEeqnarray}
Thus, by the fact $q_s$ must be an integer,
\begin{IEEEeqnarray}{c}
q_s\geq \Big\lceil \frac{w}{\lfloor n\mu\rfloor -s}\Big\rceil,
\end{IEEEeqnarray}
which implies that $\bm c^*$ is superior to $\bm c$.


\section{Adaptive Gradient Coding} \label{sec:scheme}

In this section, the achievability of Theorem \ref{theorem:basic} is proved by presenting and analyzing the AGC scheme, which achieves the optimal  communication vector $\bm c^*$ in \eqref{eqn:c:opt}.

Roughly speaking, 
	to recover a given vector from the signals of the $n-s$ active workers, we first decompose the encoding matrix into a product of two matrices in \eqref{B:decomp}, where one of them in \eqref{eqn:E} is used to satisfy some invertible properties, and the other one in \eqref{eqn:M} contains a submatrix \eqref{eqn:MU} formed by the desired vectors.
 In particular, except for the constraints of storage and decodability for $s\leq s_{\textnormal{max}}$, the column rank of first matrix is expected to be as small as possible for each decodable round, such that it can always be decoded with the minimum number of rows, i.e., the minimum communication cost.


Note that the computation load at each worker is known as ${d \over k}$, rather than the value $k$ itself. Since we are interested in achieving the optimal computation load under the memory constraint of $\mu$ in \eqref{eqn:thm}, which does not involve the value of $k$, from now on we fix $k=n$ as did in \cite{TandonLDK17,YeA18}.


From \eqref{eqn:d:size}, with $k=n$, we have
\begin{equation*}
	s+\Big\lceil {w\over q_s}\Big\rceil \leq d \leq \lfloor n\mu \rfloor.
\end{equation*}
To have the maximum straggler tolerance ability, we  choose  
\begin{eqnarray*}
d=\lfloor n\mu\rfloor.
\end{eqnarray*}
Since $k=n$, the data set $\mathcal{D}$ is equally split into $n$ equal-size subsets $\mathcal{D}_0,\mathcal{D}_1,\ldots,\mathcal{D}_{n-1}$, and each worker $j\in[0:n)$ stores $d$ subsets $\{\mathcal{D}_{j}, \mathcal{D}_{(j+1)_{(n)}},\ldots,\mathcal{D}_{(j+d-1)_{(n)}}\}$,
 where $(\cdot)_{(n)}$ is the module $n$ operation. The corresponding partial gradients $g_0,g_1,\cdots, g_{n-1}$ are defined in \eqref{eqn:gr}  and the full gradient is given by \eqref{eqn:gradient:g}.

Then, we split each partial gradient vector $g_i$ ($i\in[0:n)$) into $L$ equal dimensional sub-vectors, each of dimension\footnote{If $L\,\nmid\,w$, the gradient vectors $g_0,g_1,\ldots,g_{n-1}$ are extended to dimension $L\lceil\frac{w}{L}\rceil$ by padding some zeros to the vectors. For simplicity, we still use $g_0,\ldots,g_{n-1}$ to denote the extended vectors.} $\lceil \frac{w}{L}\rceil$, i.e.,
\begin{eqnarray*}
 g_i=(g_i(0)^\top,g_i(1)^\top,\ldots,g_i(L-1)^\top)^\top\in\mathbb{R}^{L\lceil\frac{w}{L}\rceil},
 \end{eqnarray*}
where $g_i(0),g_i(1),\ldots,g_i(L-1)\in \mathbb{R}^{\lceil {w\over L}\rceil }$, and $L$ can be any positive integer no larger than $w$. When there are $s$ stragglers present,  each active worker sends $r_s$ rounds of coded signals to the master for some $r_s\in[0:L)$, each of length $\lceil\frac{w}{L}\rceil$. Accordingly, the communication cost  is given by
 \begin{IEEEeqnarray}{c}
 {q_s\over w}={r_s\over w}\cdot\Big\lceil \frac{w}{L}\Big\rceil.\label{eqn:qs:AGC}
 \end{IEEEeqnarray}

In what follows, we describe the detailed encoding and decoding procedures for AGC code.

\textit{Encoding:} First, define
 \begin{eqnarray} \label{eqn:tilde:g}
\widetilde{g} \! \triangleq \! \left(g_0(0),\ldots,g_{n\!-\!1}(0),\ldots,g_0(\! L\!-\!1),\ldots,g_{n\!-\!1}(\! L\!-\!1)\right)^\top\!.
 \end{eqnarray}
 The AGC code is characterized by an encoding matrix $B=[B_{u,v}]_{0\le u,v<nL}$ of form:
 \begin{IEEEeqnarray}{c}
 \setlength{\arraycolsep}{0.5pt}
	B =
	\left[
	\begin{array}{c}
  		B^{(0)} 	\\
  		B^{(1)}	\\
  		\vdots	\\
  		B^{(L-1)}	\\
	\end{array}
	\right]
	=
	\left[
	\begin{matrix}
  		B(0,0) 	&		\ldots 	&	B(0,L-1) \\
 		B(1,0) 	&		\ldots 	&	B(1,L-1) \\
 		\vdots 	&		\vdots 	&	\vdots \\
 		B(L-1,0)	&	\ldots 	&	B(L-1,L-1) \\
	\end{matrix}
	\right],\IEEEeqnarraynumspace\label{eqn:B}
\end{IEEEeqnarray}
 where $B(r,m)\in\mathbb{R}^{n\times n}$ for all $r,m\in[0:L)$. Based on $B$, each worker $j\in[0:n)$ sends the signals in at most $L$ rounds sequentially. In particular, in the $r$-th round, worker $j$ sends a vector of length $\lceil\frac{w}{L}\rceil$ as
\begin{eqnarray*}
X_{j,r}\triangleq B^{(r)}_{j,:}\cdot\widetilde{g},\quad\forall\,r\in[0:L),j\in[0:n),
\end{eqnarray*}
where $B^{(r)}_{j,:}$ is the $j$-th row vector of the sub-matrix $B^{(r)}=[B(r,0),\cdots, B(r,L-1)]$.

 Recall that worker $j$ can only calculate the partial gradients $g_{j},g_{(j+1)_{(n)}},\ldots,g_{(j+d-1)_{(n)}}$. Then, each $B(r,m)$ have the following cyclic form
\begin{eqnarray}\label{eqn:Blm}
\hspace{-12mm}&&~~~\overbrace{~~~~~~~~~~~~~~~~~~~~}^{d}	\nonumber \\ 	
\hspace{-12mm}&&	
\left[
	\begin{array}{cccccccccc}
 		\ast  &		\ast & \ldots 	&	\ast		&	\ast 	&  0  &		0 & \ldots 	&	0		&	0 	\\
 		0  &	\ast 	& 	\ldots		&	\ast	&	\ast 	&  \ast  &		0 & \ldots 	&	0		&	0 	\\
 		\vdots  &		\vdots  & \vdots 	&	\vdots	&	\vdots 	&  \vdots  &	\ddots	 & \ddots 	&	\vdots		&	\vdots 	\\
 		0  &	0	 & \cdots 	&	0		&	0 	&  \ast  &	\ast	 & \ldots 	&	\ast		&	\ast 	\\
 		\vdots  &		\vdots  & \vdots 	&	\vdots	&	\vdots 	&  \vdots  &	\ddots	 & \ddots 	&	\vdots		&	\vdots 	\\
 		\ast  &		\ast & \ldots 	&	\ast		&	0 	&  0  &		0 & \ldots 	&	0		&	\ast 	
	\end{array}
	\right],
\end{eqnarray}
where $``\ast"$ represents a non-zero entry. 
Such matrix $B$ can be constructed by the product of two matrices as
\begin{eqnarray}
B=E\cdot M\label{B:decomp}
\end{eqnarray}
where the matrices $E\in\mathbb{R}^{nL\times (n-d+1)L}$ and $M\in \mathbb{R}^{(n-d+1)L\times nL}$ are generated as follows.
\begin{itemize}
  \item The matrix $E$ is given by
    \begin{eqnarray}
	E =\left[
		\begin{array}{ccccccccc}
			E^{(0)} \quad  \bm{0}_{n\times (n-d)(L-1)}	\\
			E^{(1)} \quad \bm{0}_{n\times (n-d)(L-2)} 		\\
			\vdots	\\
			E^{(L-2)}\quad \bm{0}_{n\times (n-d)}		\\
			E^{(L-1)}  		\\
			\end{array}
		\right],\label{eqn:E}
\end{eqnarray}
where $E^{(r)}$ ($r\in [0:L)$) is an $n\times (L+(r+1)(n-d))$ matrix, and all the entries in $E^{(r)}$ are randomly and independently drawn from a continuous probability distribution (e.g., Gaussian distribution).

  \item The matrix $M$ has the form
  \begin{eqnarray} \label{eqn:M}
  M=\left[\begin{matrix}
            M^U \\
            M^D
          \end{matrix}
  \right],
  \end{eqnarray}
  where $M^U\in\mathbb{R}^{L\times nL}$  is 
  given by
  \begin{eqnarray} \label{eqn:MU}
	M^U = \left[
		\begin{array}{ccccc}
			\bm{1}_{1\times n}	&	\bm{0}_{1\times n}	&	\cdots	&	\bm{0}_{1\times n}			\\
			\bm{0}_{1\times n}			&	\bm{1}_{1\times n} &	\ldots	&	\bm{0}_{1\times n}			\\
			\vdots	&	\vdots	&	\ddots	&	\vdots	\\
			\bm{0}_{1\times n}			&	\bm{0}_{1\times n}	&		\ldots	&	\bm{1}_{1\times n}			
		\end{array}
		\right],	
\end{eqnarray}
and $M^D\in\mathbb{R}^{L(n-d)\times nL}$ is specified such that all the sub-matrix $B(r,m)$ has the form as in \eqref{eqn:Blm}.
In fact, $M^D$  can be determined uniquely as below.
For each $v\in[0:nL)$, let
\begin{eqnarray}\label{eqn:Q_v}
 \mathcal{Q}_v\triangleq\{u\in[0:nL): B_{u,v}=0\},
 \end{eqnarray}
 i.e., the set of row indices in the $v$-th column of $B$ with deterministic zero entries.  By \eqref{eqn:B} and \eqref{eqn:Blm}, $|\mathcal{Q}_v|=(n-d)L$. Then, the matrix $M^D$ satisfies
 \begin{eqnarray} \label{eqn:MuMd}
   [E_{\mathcal{Q}_v,[0:L)}~E_{\mathcal{Q}_v,[L:(n\!-\!d\!+\!1)L)}] \! \cdot \!
   \left[
   \begin{matrix}
            M^U_{:,v}  \\
            M^D_{:,v}
    \end{matrix}
   \right]
   \! =\! \bm{0}_{(n-d)L\times 1}.\!
 \end{eqnarray}
It is seen from \eqref{eqn:Blm} and \eqref{eqn:Q_v} that $\mathcal{Q}_v$ contains $(n-d)$ row indices in each $B^{(r)}, r\in [0:L)$. That is, we can express $\mathcal{Q}_v$ as $\mathcal{Q}_v = \{\mathcal{Q}_v^{(r)},r\in[0:L)\}$ where $\mathcal{Q}_v^{(r)} \triangleq\{u\in[0:n): B^{(r)}_{u,v}=0\}$ and $|\mathcal{Q}_v^{(r)}|=n-d$.
Consequently, the matrix $E_{\mathcal{Q}_v,:}$ has the following structure
\begin{eqnarray*}
	E_{\mathcal{Q}_v,:}=\left[
		\begin{array}{ccccccccc}
			E^{(0)}_{\mathcal{Q}_v^{(0)},:} \quad  \bm{0}_{(n-d)\times (n-d)(L-1)}	\\
			E^{(1)}_{\mathcal{Q}_v^{(1)},:} \quad \bm{0}_{(n-d)\times (n-d)(L-2)} 		\\
			\vdots	\\
			E^{(L-2)}_{\mathcal{Q}_v^{(L-2)},:}\quad \bm{0}_{(n-d)\times (n-d)}		\\
			E^{(L-1)}_{\mathcal{Q}_v^{(L-1)},:}  		\\
			\end{array}
		\right],
\end{eqnarray*}
where $E^{(r)}_{\mathcal{Q}_v^{(r)},:}$ is an $(n-d) \times (L+(r+1)(n-d))$ matrix.
Then,  $E_{\mathcal{Q}_v, [L:(n-d+1)L )}$ is indeed a lower triangular block matrix of the form
\begin{eqnarray}
\setlength{\arraycolsep}{0.5pt}
\left[\begin{array}{cccc}
  (*)_{(n-d)\times(n-d)} & \bm{0}_{(n-d)\times(n-d)} & ... & \bm{0}_{(n-d)\times(n-d)} \\
  (*)_{(n-d)\times(n-d)}&(*)_{(n-d)\times(n-d)}&...& \bm{0}_{(n-d)\times(n-d)}\\
  \vdots&\vdots&\ddots&\vdots\\
  (*)_{(n-d)\times(n-d)}&(*)_{(n-d)\times(n-d)}&... &(*)_{(n-d)\times(n-d)}
\end{array}\right],
\label{eqn:E_Qv}
\end{eqnarray}
where $(*)_{(n-d)\times(n-d)}$ denotes a matrix of size $(n-d)\times (n-d)$ with all entries randomly drawn from a continuous distribution. Obviously, $\det(E_{\mathcal{Q}_v,[L:(n-d+1)L)})$ can be seen as a non-zero polynomial of the entries of $E_{\mathcal{Q}_v,[L:(n-d+1)L)}$, and thus $\{E_{\mathcal{Q}_v,[L:(n-d+1)L)}|\det(E_{\mathcal{Q}_v,[L:(n-d+1)L)})=0\}$ is a set measure zero (also referred to null set in measure theory). Then, we have $\Pr\{\det(E_{\mathcal{Q}_v,[L:(n-d+1)L)})=0\}=0$, i.e., the square matrix $E_{\mathcal{Q}_v,[L:(n-d+1)L)}$ is invertible with probability $1$. Therefore, by \eqref{eqn:MuMd}, for any  $v\in[0:nL)$, the $v$-th column of $M^D$ is determined by
\begin{eqnarray*}
 M^D_{:,v}
= -( E_{\mathcal{Q}_v,[L:(n\!-\!d\!+\!1)L)})^{-1}\! \cdot \! E_{\mathcal{Q}_v,  [0:L)} \! \cdot \! M^U_{:,v}.
\end{eqnarray*}
 \end{itemize}


\textit{Decoding:}
Given $s\in[0:d)$, let
\begin{IEEEeqnarray}{c}
r_s=\Big\lceil \frac{L}{d-s}\Big\rceil.\label{eqn:ls}
\end{IEEEeqnarray}
We show that when there are $s$ stragglers present,  the master can decode the full gradient \eqref{eqn:gradient:g} from $\{X_{j,r}:j\in\mathcal{F},r\in[0:r_s)\}$ for
  any  active workers set  $\mathcal{F}$ with $|\mathcal{F}|=n-s$.

The received signal at the master is
\begin{eqnarray*}
	E^{(\mathcal{F})}\cdot M\cdot \widetilde{g}	,	
\end{eqnarray*}
where $E^{(\mathcal{F})}$ is a matrix of size ${(n-s)r_s\times L(n-d+1)}$ given by
\begin{eqnarray} \label{eqn:EH}
	E^{(\mathcal{F})} \triangleq \left[
		\begin{array}{ll}
			E^{(0)}_{\mathcal{F},:}	&	\textbf{0}_{(n-s)\times (L-1)(n-d)}	\\
			~~~\vdots &~~~	\vdots	\\
			E^{(r_s-2)}_{\mathcal{F},:} 	&	\textbf{0}_{(n-s)\times (L-r_s+1)(n-d)}	\\
			E^{(r_s-1)}_{\mathcal{F},:} &	\textbf{0}_{(n-s)\times(L-r_s)(n-d)}	\\
		\end{array}
		\right],
\end{eqnarray}
where the entries of the $(n-s)\times (L+(r+1)(n-d))$ matrix $E^{(r)}_{\mathcal{F},:}$ are randomly and independently generated according to some continuous distribution for all $r\in[0:r_s)$.
We would decode $g$ with the first $L+(n-d)r_s$ signals of $E^{(\mathcal{F})}\cdot M\cdot \widetilde{g}$. Let $\mathcal{H}=[0:L+(n-d)r_s)$.  Since the columns of $E^{(\mathcal{F})}$ in \eqref{eqn:EH}  are all zero vectors except for the first $L+(n-d)r_s$ ones, we have
\begin{eqnarray}\label{eqn:encode_emg}
E^{(\mathcal{F})}_{\mathcal{H},:}\cdot M\cdot \widetilde{g}=E_{\mathcal{H},\mathcal{H}}^{(\mathcal{F})}\cdot M_{\mathcal{H},:}\cdot\widetilde{g}.\label{eqn:signal:a}
\end{eqnarray}
Given that $E_{\mathcal{H},\mathcal{H}}^{(\mathcal{F})}$ is invertible, we can decode $M_{\mathcal{H},:}\cdot\widetilde{g}$ by multiplying the matrix $\big(E_{\mathcal{H},\mathcal{H}}^{(\mathcal{F})}\big)^{-1}$ to \eqref{eqn:encode_emg}. Note from \eqref{eqn:M} and \eqref{eqn:MU}, the first $L$ rows of $M_{\mathcal{H},:}\cdot\widetilde{g}$ form the vector $\sum_{i=0}^{n-1}g_i$. That is, we are able to decode the full gradient successfully.

In the following, we complete the proof by  proving that $E_{\mathcal{H},\mathcal{H}}^{(\mathcal{F})}$ is invertible with probability $1$, i.e., $\Pr\{\det(E_{\mathcal{H},\mathcal{H}}^{(\mathcal{F})})\neq 0\}=1$.
%

We first show that the entries in the diagonal of the matrix $E_{\mathcal{H},\mathcal{H}}^{(\mathcal{F})}$ are randomly generated non-zero values. To see that, for any $i\in [0:L+(n-d)r_s )$, let $i=i_1(n-s)+i_2$ where $i_1\in[0: r_s)$ and $i_2 \in[0:n-s)$. Then, the entry of $E_{\mathcal{H},\mathcal{H}}^{(\mathcal{F})}$ at row $i$ and column $i$ is an entry of the $i_1$-{th} block matrix $E ^{(i_1)}_{\mathcal{F},:}$ given in \eqref{eqn:EH} since $E_{\mathcal{F},:}^{(i_1)}$ has $L+(i_1+1)(n-d)$ columns and
\begin{itemize}
\item If $0\le i_1 < r_s-1$
\begin{eqnarray*}
i&=&i_1(n-s)+i_2\\
&<& (i_1+1)(n-s)\\
&< & L+(i_1+1)(n-d)
\end{eqnarray*}
where the last inequality follows from $r_s=\lceil\frac{L}{d-s}\rceil<\frac{L}{d-s}+1$ and then $(i_1+1)(d-s)\le (r_s-1)(d-s)< L$.
\item  If $i_1=r_s-1$, then $i< L+(n-d)r_s=L+(i_1+1)(n-d)$ due to the fact $i\in[0:L+(n-d)r_s )$.
\end{itemize}
Therefore, $\det(E_{\mathcal{H},\mathcal{H}}^{(\mathcal{F})})$ is a non-zero polynomial containing the product of the elements in the diagonal of $E_{\mathcal{H},\mathcal{H}}^{(\mathcal{F})}$. Then, $\{\det(E_{\mathcal{H},\mathcal{H}}^{(\mathcal{F})})|\det(E_{\mathcal{H},\mathcal{H}}^{(\mathcal{F})})=0\}$ is a zero measure set and $\Pr\{\det(E_{\mathcal{H},\mathcal{H}}^{(\mathcal{F})})=0\}=0$.  This completes the proof of achievability.

\textit{Performance:} In the above scheme, when there are $s$ stragglers present, the communication cost is given by \eqref{eqn:qs:AGC}, where $r_s$ is given by \eqref{eqn:ls} and $d=\lfloor n\mu\rfloor$. Thus, the scheme achieves the communication vector
\begin{eqnarray}
\bm c=\frac{1}{w}\bigg(\Big\lceil \frac{L}{\lfloor n\mu\rfloor}\Big\rceil\!\cdot\!\Big\lceil\frac{w}{L}\Big\rceil,\Big\lceil \frac{L}{\lfloor n\mu\rfloor\!-\!1}\Big\rceil\!\cdot\!\Big\lceil\frac{w}{L}\Big\rceil,\ldots, L\!\cdot\!\Big\lceil\frac{w}{L}\Big\rceil \bigg),\label{eqn:AGC:c}
\end{eqnarray}
where $L$ can be any positive integer no larger than $w$.
In fact, there is a tradeoff between $L$ and the communication cost. For instance, setting $L=1$, $\bm c$ in \eqref{eqn:AGC:c} will degenerate to the performance of maximum communication cost 1 for any real-time number of stragglers $s\leq d-1$, while setting $L=w$ in \eqref{eqn:AGC:c},  the achievability of $\bm c^*$ is obtained.

\paragraph*{Complexity} 
The decoding complexity is dominated by the complexity of inverting the encoding matrix of AGC/CGC. For the random matrix used in AGC, the complexity is bounded by $O((\cdot)^3)$,  while for Vandermonde matrix used in CGC, it can be reduced to $O((\cdot)^2)$ \cite{DemeureS89}, where $(\cdot)$ is the size of matrix.
 In the presence of  $s$ stragglers,  if $L$ is chosen to be $w$ to achieve $\bm c^*$, the matrix that AGC needs to invert is $(w+(n-d)\lceil \frac{w}{d-s}\rceil)\times (w+(n-d)\lceil\frac{w}{d-s}\rceil)$, while for CGC is $(\lceil {w\over q} \rceil+n-d)\times (\lceil {w\over q} \rceil+n-d)$.  Thus, the inverting complexity of AGC  $O((w+(n-d)\lceil \frac{w}{d-s}\rceil)^3)$ is larger than CGC with $O((\lceil {w\over q} \rceil+n-d)^2)$ since $\lceil {w\over q} \rceil \leq w$ and $\lceil\frac{w}{d-s}\rceil \geq 1$.

However, to reduce the complexity of AGC for sufficient large $w$, we can take $L= \mathrm{lcm} (1,2,\cdots,d)$ to approximately approach $\bm c^*$ since in the case of   $\mathrm{lcm} (1,2,\cdots,d)|w$,
\begin{itemize}
\item the ceiling operations in \eqref{eqn:c:opt} can be removed, which results in $\bm {c}^{\ast} = ({1\over \lfloor n\mu\rfloor},{1\over \lfloor n\mu\rfloor-1},\cdots,1)$;
\item  by setting  $L= \mathrm{lcm} (1,2,\cdots,d)$, the  ceiling operations in \eqref{eqn:AGC:c} can be removed, which results in $\bm {c} = ({1\over \lfloor n\mu\rfloor},{1\over \lfloor n\mu\rfloor-1},\ldots,1)=\bm{c}^\ast$. 
\end{itemize}
As $w$ is typically very large and $d$ is relatively small, $\mathrm{lcm} (1,2,\cdots,d)|w$ can always be satisfied by padding a few zeros at the end of the gradient vectors, which only results in little performance loss. 
In this case, we only need to invert matrix $E_{\mathcal{H},\mathcal{H}}^{(\mathcal{F})}$ of size $(L+(n-d)\lceil \frac{L}{d-s}\rceil)\times (L+(n-d)\lceil\frac{L}{d-s}\rceil)$ instead of $(w+(n-d)\lceil\frac{w}{d-s}\rceil)\times (w+(n-d)\lceil\frac{w}{d-s}\rceil)$, which results in the complexity of $O( n^3)$ when $d,L$ are small fixed values satisfying $d\leq L \ll n$.
It is verified in Section \ref{sec:amazon} that, the overall average running time of AGC is smaller than that of CGC due to its advantage in communication cost. Therefore, the complexity introduced by AGC is acceptable. 

\paragraph*{Numerical stability}
Normally, the numerical stability is determined by the condition number\footnote{Here we only consider the condition number for inversion.} of the matrix $E_{\mathcal{H},\mathcal{H}}^{(\mathcal{F})}$ in (29), 
which is used to measure how sensitive a function is to changes or errors in the input, and how much error in the output results from an error in the input. However, due to the irregular form of the  matrix $E_{\mathcal{H},\mathcal{H}}^{(\mathcal{F})}$, it can not be calculated by directly using the existing works  as did in \cite{YeA18,Kadhe20}. 
Therefore, we give the following theorem to demonstrate the relationship of the condition number in AGC with large enough $n$, whose proof can be found in the full version of our paper \cite{Cao21} because of the space limitation here.
\begin{Theorem}\label{thm:numerical}
	Consider a $(n,\mu,w)$ distributed system, when $n$ is large, the 2-norm condition number in the decoding phase of AGC scheme is in order $O(n^{1\over 6\epsilon})$ with probability at least $1-\epsilon$ for any  $0<\epsilon<1$. Particularly, when $\epsilon=1/6$, there is at least $1-\epsilon\approx 83\%$ probability that the condition number is bounded by some linear function of $n$.
\end{Theorem}
\begin{Remark}
	Note that, although the probability is not equal to 1, the numerical stability that linearly increase with $n$ can still be guaranteed by searching for multiple realizations of random matrix with computer. This cost is affordable since the construction of encoding matrix is only one-time.
\end{Remark}

\begin{Remark}(A Conjecture on Structured Construction)
A heuristic construction of $E$ in \eqref{eqn:E} is to start with  Vandermonde-like matrix, and then replace the elements at the zero-positions indicated by \eqref{eqn:E} by zero. That is, let $\{\alpha_{i,r}:i\in[0:n),r\in[0:L)\}$ be $nL$ distinct non-zero elements over real field, then for each $r\in[0:L)$, the submatrix $E^{(r)}$ in \eqref{eqn:E} is constructed based on: 
\begin{IEEEeqnarray}{c}
E^{(r)}=\left[\begin{array}{ccccc}
1&\alpha_{0,r}&\alpha_{0,r}^2&\ldots&\alpha_{0,r}^{L+(r+1)(n-d)-1}\\
1&\alpha_{1,r}&\alpha_{1,r}^2&\ldots&\alpha_{1,r}^{L+(r+1)(n-d)-1}\\
\vdots&\vdots&\vdots&\ddots&\vdots\\
1&\alpha_{n-1,r}&\alpha_{n-1,r}^2&\ldots&\alpha_{n-1,r}^{L+(r+1)(n-d)-1}
\end{array}\right].\notag
\end{IEEEeqnarray}
We conducted extensive experiments, and verified that such matrices satisfy our decoding conditions, i.e., 
\begin{itemize}
\item $E_{\mathcal{Q}_v, [L:(n-d+1)L )}$ in \eqref{eqn:E_Qv} is of full rank for each $v\in[0:nL)$;
\item $E_{\mathcal{H},\mathcal{H}}^{(\mathcal{F})}$ in \eqref{eqn:signal:a} is invertible for all $\mathcal{F}\subseteq[0:n)$ with $|\mathcal{F}|=n-s$ for each $s\in[0:d)$. 
\end{itemize}
However, due to the irregular form  of $E$ in \eqref{eqn:E}, it is hard to verify the above conditons mathematically. We leave this as an open problem for future exploration. 
\end{Remark}

\begin{Remark}[Technique difference compared to CGC]
In fact, our work is inspired by the work CGC of \cite{YeA18}. The difference is, the key technique in our paper is to redesign a new encoding matrix $E$ to create a lower block-triangular structure, such that the column rank of the received matrix in the master can be varied with the number of communication rounds. This is achieved by padding zeros in some certain places to the encoding matrix $E$ and allowing workers to send multiple rounds of signals. Further, such structure makes it possible to have various invertible size, thus the system can adaptively decode the desired messages in the lowest communication cost according to the real-time number of stragglers. 
 While the encoding matrix in [19] is a regular Vandermonde matrix, whose invertible size is $n-s_{\max}$ for some fixed $s_{\max}$ and it can not be directly extended to cope with the varying number of stragglers adaptively in the practical applications.
\end{Remark}

\begin{Example}
Consider the encoding matrix of AGC in Example \ref{eg:motivation}, it can be constructed by $B =
E\cdot M$ as
\begin{eqnarray*}
\left[
 \begin{matrix}
3	&	2	&	1	&	0		\\
3	&	1	&	1	&	0		\\
1	&	3	&	2	&	0		\\	
2	&	1	&	3	&	3		\\
2	&	3	&	2	&	3		\\
2	&	1	&	1	&	3		\\
\end{matrix}
\right]
\cdot
\left[
 \begin{matrix}
1	&	1	&	1	&	0	&	0	&	0	\\
0	&	0	&	0	&	1	&	1	&	1	\\
-3	&	-{1\over 2}	&	-3	&	-1	&	-{3\over 2}	&	-2	\\
{4\over 3}	&	-{1\over 2}	&	{7\over 3}	&	-{1\over 3}	&	{1\over 6}	&	{5\over 3}	\\
\end{matrix}
\right].
\end{eqnarray*}

As for decoding, when there is no straggler, after receiving the first three messages, the master will broadcast a stop signal to workers. Then by multiplying $(E_{[0:3),[0:3)}^{(\{0,1,2\})})^{-1}$ with $E_{[0:3),[0:3)}^{(\{0,1,2\})}\cdot M \cdot \widetilde{g}$ where $E_{[0:3),[0:3)}^{(\{0,1,2\})}=E_{[0:3),[0:3)}$, the master can decode:
\begin{eqnarray*}
\setlength{\arraycolsep}{0.5pt}
\left[
 \begin{matrix}
1	&	1	&	1	&	0	&	0	&	0	\\
0	&	0	&	0	&	1	&	1	&	1	\\
-3	&	-{1\over 2}	&	-3	&	-1	&	-{3\over 2}	&	-2	\\
\end{matrix}
\right]
\cdot
\left[
 \begin{matrix}
g_0(0)	\\
g_1(0)	\\
g_2(0)	\\
g_0(1)	\\
g_1(1)	\\
g_2(1)	\\
\end{matrix}
\right]	
=
\left[
 \begin{matrix}
\sum_{j=0}^{2}g_j(0)	\\
\sum_{j=0}^{2}g_j(1)	\\
\ast
\end{matrix}
\right],
\end{eqnarray*}
where the first two rows  form the desired full gradient, and $\ast$ represents useless coordinate. When there exists one straggler, for example we assume worker $2$ is a straggler, after receiving four messages from the active workers, the master broadcasts stop message. Then by multiplying $(E_{\{0,1,3,4\},\{0,1,3,4\}}^{(\{0,1\})})^{-1}$ with $E_{\{0,1,3,4\},\{0,1,3,4\}}^{(\{0,1\})} \cdot M \cdot \widetilde{g}$, the master can decode the full gradient from the first two rows as follows:
\begin{eqnarray*}
\setlength{\arraycolsep}{0.5pt}
\left[
 \begin{matrix}
1	&	1	&	1	&	0	&	0	&	0	\\
0	&	0	&	0	&	1	&	1	&	1	\\
-3	&	-{1\over 2}	&	-3	&	-1	&	-{3\over 2}	&	-2	\\
{4\over 3}	&	-{1\over 2}	&	{7\over 3}	&	-{1\over 3}	&	{1\over 6}	&	{5\over 3}	\\
\end{matrix}
\right]
\cdot
\left[
 \begin{matrix}
g_0(0)	\\
g_1(0)	\\
g_2(0)	\\
g_0(1)	\\
g_1(1)	\\
g_2(1)	\\
\end{matrix}
\right]
=
\left[
 \begin{matrix}
\sum_{j=0}^{2}g_j(0)	\\
\sum_{j=0}^{2}g_j(1)	\\
\ast	\\
\ast
\end{matrix}
\right].
\end{eqnarray*}
\end{Example}

\section{Group AGC Scheme}\label{sec:Group}

For large distributed computing systems, as demonstrated in \cite{TandonLDK17,YeA18,Kadhe20}, large $n$ usually incorporates high complexity in encoding/decoding procedures. An approach to decrease the complexity is partitioning the $k$ data subsets and $n$ workers into multiple smaller groups respectively, where the workers in each group form a smaller subsystem that is responsible for computing the partial gradients over the data subsets assigned to it. This method can also provide stronger robustness in the sense that it can  tolerate more stragglers in certain conditions. In this section,  we incorporate this idea into AGC, which will be referred to as G-AGC scheme. We start with an illustrative example.

\begin{Example}\label{exg:group}
	Consider a $(7,{2\over 7},2)$ system, let $k=n=7$. The data subsets $\{\mathcal{D}_i\}_{i=0}^6$ and workers $[0:7)$ are partitioned into three groups $\{\mathcal{D}_0,\mathcal{D}_1\}, \{\mathcal{D}_2,\mathcal{D}_3\}, \{\mathcal{D}_4,\mathcal{D}_5,\mathcal{D}_6\} $ and $\{0,1\},\{2,3\},\{4,5,6\}$ respectively. The three groups of workers are responsible for computing the partial gradients over the three groups of data subsets respectively. The master decodes the full gradient by adding the partial gradients computed by the subsystems, as illustrated in Fig. \ref{fig:G-AGC}. Notice that the three groups of workers form $(2,1,2)$, $(2,1,2)$ and $(3,\frac{2}{3},2)$ systems respectively. The AGC scheme with the choice of $L=2$ is applied to the three subsystems respectively. According to Theorem \ref{theorem:basic}, the communication vector $\bm c=(\frac{1}{2},1)$  is achieved for all the three subsystems, and thus is also achievable for the whole system.

Denote the number of stragglers in the three subsystems by $s_0, s_1$ and $s_2$ respectively, then G-AGC can successfully compute the full gradient when all the subsystems can compute their partial gradients, i.e., $s_j<2$ for all $j\in\{0,1,2\}$.
We observe that, even if $s_0=s_1=s_2=1$, G-AGC can still compute the full gradient at the communication cost of $1$ for each active worker, while the $(7,{2\over 7},2)$ system using AGC directly can only tolerate at most 1 straggler. Therefore, G-AGC scheme is more robust to the stragglers compared to AGC in some cases.

\begin{figure*}[!htb]
\setlength{\abovecaptionskip}{-0.1cm} 	
	\centering
		\includegraphics[width=0.8\textwidth]{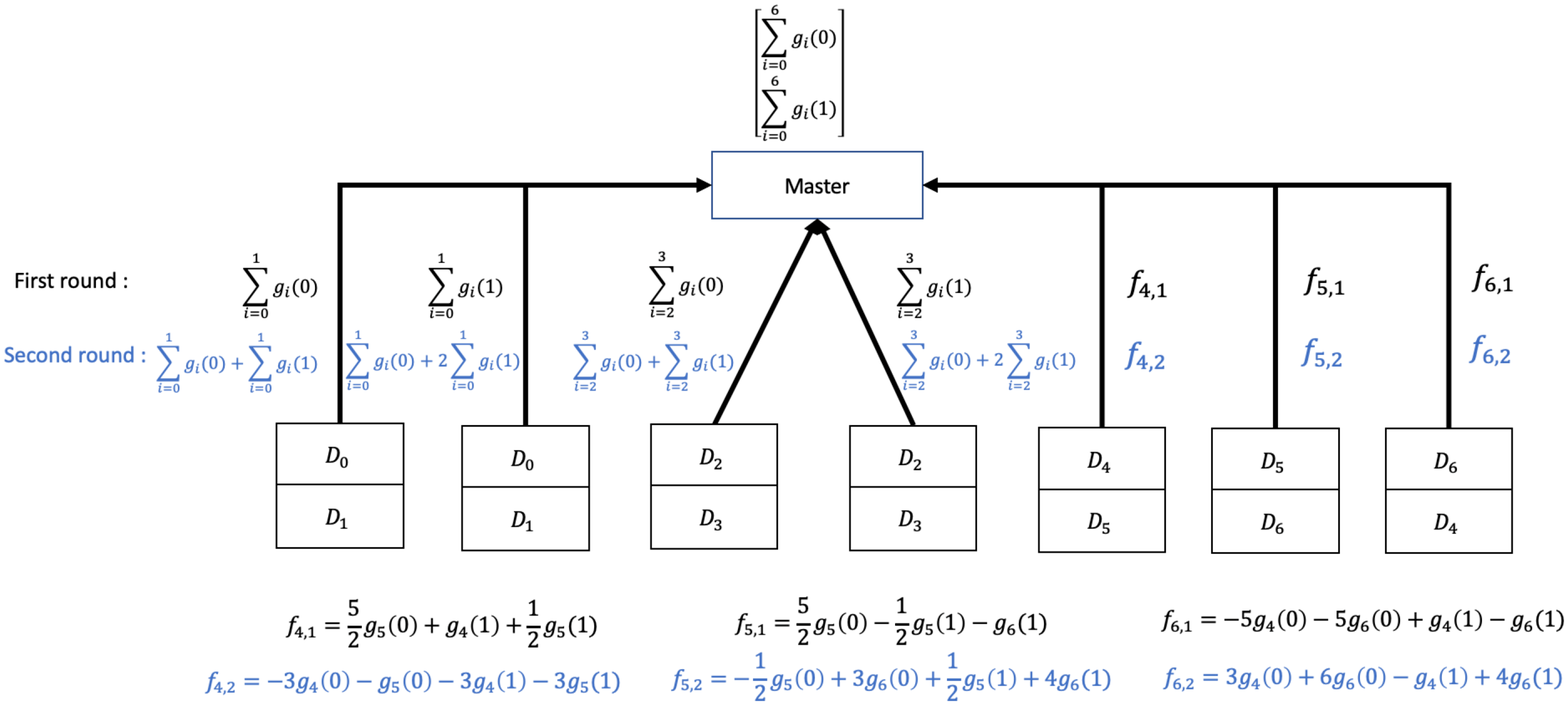}
\caption{Illustration of G-AGC scheme for a $(n,\mu,w)=(7,\frac{2}{7},2)$ system, with the choice of $L=2$.}\label{fig:G-AGC}	
\end{figure*}
\end{Example}

In the following, we extend the above example to general case.

For a given $(n,\mu,w)$ system, let $d=\lfloor n\mu\rfloor $, $k=n$, i.e., the data $\mathcal{D}$ is equally partitioned into $n$ data subsets $\mathcal{D}_0,\ldots,\mathcal{D}_{n-1}$, and each worker stores $d=\lfloor n\mu\rfloor$ subsets. Partition the indices $[0:n)$ into $m+1$ subgroups $\mathcal{I}_0,\mathcal{I}_1,\ldots,\mathcal{I}_{m}$, where $m\triangleq \lfloor \frac{n}{d}\rfloor -1$ and
\begin{IEEEeqnarray}{c}
\mathcal{I}_j=\left\{\begin{array}{ll}
                       [jd:(j+1)d), &\textnormal{if}~j=0,\ldots,m-1  \\
                     \mbox{$[jd:n),$}   &  \textnormal{if}~j=m
                     \end{array}
\right.,\label{eqn:Ij}
\end{IEEEeqnarray}
For each $j\in [0:m]$, define the $j$-th partial gradient by
\begin{equation}
	\bar{g}^{(j)} = \sum_{i\in\mathcal{I}_j}g_{i}.\label{partial:gradient}
\end{equation}
The $j$-th subsystem, consisting of the workers in $\mathcal{I}_j$, is   responsible for computing the partial gradient \eqref{partial:gradient}. Thus, those workers with data subsets $\{\mathcal{D}_i:i\in\mathcal{I}_j\}$ form a
       \begin{itemize}
         \item  $(d,1,w)$ system if $j=0,1,\ldots,m-1$;
         \item  $\big(n-md,\frac{d}{n-md},w\big)$ system if $j=m$.
       \end{itemize}
       
The encoding and decoding processes are as follows.

\paragraph*{Encoding}
Choose some parameter $L\in[1:w]$. For each  $j\in[0:m]$, the workers in the subsystem $j$ encode the gradients according to the encoding phase in Section \ref{sec:scheme} with its data subsets  $\{\mathcal{D}_i:i\in\mathcal{I}_j\}$ and parameters of subsystem $j$.

\paragraph*{Decoding}
Let $s_j$ be the number of stragglers in the $j$-th subsystem. For the $j$-th subsystem, the master can decode the partial gradient $\bar{g}^{(j)}$ in \eqref{partial:gradient} according to the decoding phase in Section \ref{sec:scheme} after receiving $r_{j}=\lceil\frac{L}{d- s_j}\rceil$ signals from the active workers. Since subsystems work in parallel, and the master needs to collect all the $m+1$ partial gradients $\overline{g}^{(0)},\overline{g}^{(1)},\ldots,\overline{g}^{(m)}$,  the communication length (see \eqref{eqn:signal}) is given by
\begin{IEEEeqnarray}{rcl}
q_{(s_0,\ldots,s_m)} &\triangleq& \max_{j\in[0:m]}r_{j}\!\cdot \! \Big\lceil\frac{w}{L}\Big\rceil
= \Big\lceil{L\over d \!-\! \max\{s_0,\!\ldots \!,s_m\}}\Big\rceil\cdot \Big\lceil\frac{w}{L}\Big\rceil.\label{eqn:com:length}
\end{IEEEeqnarray}
Then, the master computes the full gradient \eqref{eqn:gradient:g} by adding up all the partial gradients in \eqref{partial:gradient}, i.e.,
       \begin{eqnarray}
       g=\sum_{j=0}^{m}\bar{g}^{(j)} = \sum_{j=1}^{m}\sum_{i\in\mathcal{I}_j}g_{i}=\sum_{i=0}^{n-1}g_i.\label{eqn:group:g}
       \end{eqnarray}

\begin{Remark}[Group versions of CGC]\label{remark:gcgc}
In fact, the idea of group has been  incorporated into CGC in \cite{Kadhe20}. It is worth mentioning that, following the partition for data and workers in \eqref{eqn:Ij}, one can always obtain the full gradient given the partial gradients $\overline{g}^{(1)},\ldots,\overline{g}^{(m)}$ from \eqref{eqn:group:g}. Thus, we can simply replace AGC in each subsystem with CGC schemes to obtain the group version of them, which is referred to as G-CGC. In contrast to the previous versions in \cite{Kadhe20}, which only works whenever $(d=\lfloor n\mu\rfloor)|\,n$, the G-CGC in our framework works for all positive integers $d,n$ with $d\leq n$.  Notice that, in each subsystem of the group version of CGC, the achieved communication vector is the same as that achieved without group strategy. Similar to the advantage in resisting stragglers as discussed above, the grouping version of CGC is more robust to stragglers than the original one. 
\end{Remark}

\textit{Performance:} 
%
The following theorem implies that G-AGC scheme achieves the optimal communication length/cost among all the distributed gradient computing schemes employing the same grouping strategy for any given number of stragglers $s_0,\ldots,s_m$ in the groups $0,1,\ldots,m$ respectively.
\begin{Theorem}\label{cor:G-AGC}
	Consider a $(n,\mu,w)$ distributed gradient computing system, which computes the full gradient with $m+1=\big\lfloor\frac{n}{\lfloor n\mu\rfloor}\big\rfloor$   subsystems as decomposed  in \eqref{eqn:Ij}, and decodes the full gradient from the partial gradients  \eqref{eqn:group:g}. In the case that there are $s_j< \lfloor n\mu \rfloor$ stragglers for all $j\in[0:m]$, a communication length $q_{(s_0,\ldots,s_m)}^G>0$ can be achieved if and only if
\begin{IEEEeqnarray}{c}\label{eqn:group:thm}
\lfloor n\mu\rfloor\geq \max\{s_0,\ldots,s_m\} + \Big\lceil\frac{w}{q_{(s_0,\ldots,s_m)}^G}\Big\rceil.\label{eqn:thm:group}
\end{IEEEeqnarray}
\end{Theorem}

\begin{IEEEproof} This theorem holds since 
\begin{enumerate}
  \item Assume that $q_{(s_0,\ldots,s_m)}^G$ satisfies \eqref{eqn:thm:group}, then $q_{(s_0,\ldots,s_m)}^G\geq\frac{w}{\lfloor n\mu \rfloor-\max\{s_0,\ldots,s_m\}}$. By the fact that $q_{(s_0,\ldots,s_m)}^G$ is an integer,
      \begin{IEEEeqnarray*}{c}
      q_{(s_0,\ldots,s_m)}^G\geq\Big\lceil\frac{w}{\lfloor n\mu \rfloor -\max\{s_0,\ldots,s_m\}}\Big\rceil,
      \end{IEEEeqnarray*}
      where the optimal communication length on the R.H.S can be achieved by G-AGC with $L=w$ in \eqref{eqn:com:length}.
  \item Assume that the communication length $q_{(s_0,\ldots,s_m)}^G$ is achievable. By the converse arguments in Section \ref{subsec:converse} (see \eqref{eqn:d:size}), the $j$-th subsystem must satisfy
 \begin{IEEEeqnarray*}{c}
  s_j+\Big\lceil \frac{w}{q_{(s_0,\ldots,s_m)}^G}\Big\rceil\leq \lfloor n\mu\rfloor.
  \end{IEEEeqnarray*}
 Thus, it follows that $q_{(s_0,\ldots,s_m)}^G$ must satisfy \eqref{eqn:thm:group}.
\end{enumerate}
\end{IEEEproof}

Recall Theorem \ref{theorem:basic} indicates that the scheme must satisfy $s\leq \lfloor n\mu\rfloor-1$ for any subset of stragglers with size of $s$. While the constraint in \eqref{eqn:thm:group} shows that in addition to tolerate any subset of $s\leq \lfloor n\mu\rfloor-1$ stragglers, the schemes based on group can tolerate stragglers as long as $\max\{s_0,\cdots,s_{m}\}\leq \lfloor n\mu\rfloor-1$, thus the total number of stragglers $s=\sum_{i=0}^m s_i$ may exceed $\lfloor n\mu\rfloor-1$. Theorem \ref{cor:G-AGC} indicates that G-AGC is more robust to the stragglers than AGC.

In addition, the following theorem indicates that the total number of stragglers   that the system can tolerate across the groups of the proposed G-AGC is almost optimal among all group schemes satisfying the following conditions:  
\begin{enumerate}
\item[C1] Each group computes distinct data subsets;
\item[C2] Each data subset is computed $d$ times by different workers; 
\item[C3] Each worker computes the same number of data subsets.
\end{enumerate}
The proof is deferred to the Appendix.

\begin{Theorem}\label{thm:group_lb}
	Consider a $(n,\mu,w)$ distributed gradient computing system, which partitions the workers into multiple groups. Denote $s^{\ast}_{\rm{total}}$  the optimal total number of stragglers that the system can tolerate across the groups among all group schemes satisfying the conditions C1--C3, then the  total number of stragglers that the system can tolerate across the groups  $s_{\rm{total}}^{\rm{G\text{-}AGC}}$ achieved by G-AGC satisfies: 

	\begin{equation*}
		s_{\rm{total}}^{\rm{G\text{-}AGC}} >s^{\ast}_{\rm{total}}- d+1.
	\end{equation*}
	In particular, when $d|n$ or $n=\theta d+1, \theta\geq1$, 
	\begin{equation*}
		 s_{\rm{total}}^{\rm{G\text{-}AGC}}=s^{\ast}_{\rm{total}}.
	\end{equation*}
\end{Theorem}

Moreover, for the $j$-th subsystem where $j \in[0:m)$, each worker can store all the data subsets that it needs to compute the $j$-th partial gradient $\overline{g}^{(j)}$.
This corresponds to the case $\mu=1$ and $n=d$, where the gradient to be computed in this subsystem is $\bar{g}^{(j)}$. Recall that the encoding matrix $B$ is constructed by $E\cdot M$ as in \eqref{B:decomp}, where for $n=d$,
 \begin{itemize}
   \item all entries of the matrix $E\in\mathbb{R}^{dL\times L}$ are drawn from some continuous distribution;
   \item the matrix $M=M^U\in \mathbb{R}^{L\times dL}$ is given by \eqref{eqn:MU}, and the matrix $M_D$ does not exist.
 \end{itemize}
Alternatively, the matrix $E$ can be constructed in a determined way by using any matrix of size $dL\times L$ such that any $L$ rows of $E$ is of full rank, e.g., the Vandermonde matrix
\begin{eqnarray}\label{eqn:Vandermonde}
	E	=
		\left[\begin{array}{cccc}
    1 & \alpha_0 & \ldots & \alpha_0^{L-1} \\
    1 & \alpha_1 & \ldots & \alpha_{1}^{L-1} \\
    \vdots & \vdots & \ddots & \vdots \\
    1 & \alpha_{dL-1} & \ldots & \alpha_{dL-1}^{L-1}.
  \end{array}\right],
\end{eqnarray}
where $\alpha_i,i\in[0:dL)$ are distinct values in $\mathbb{R}$.
The advantage of using a structured construction is that the complexity of decoding can be greatly reduced, because  square Vandermonde matrices
can be efficiently inverted \cite{DemeureS89}. Notice that, the condition number of the matrix $E$ in \eqref{eqn:Vandermonde} only depends on $L$, which does not scale with the number of workers $n$.

\paragraph*{Complexity}
	Since each group has its own decoding matrix, the complexity is composed of two parts: $\lfloor {n\over d}\rfloor-1$ groups each with  $d$ workers and one group with  $n-\lfloor {n\over d}\rfloor d+d$  workers. Therefore, in G-CGC, the decoding matrix is size of $\lceil {w\over q} \rceil \times \lceil {w\over q} \rceil$ and $(\lceil {w\over q} \rceil+n-\lfloor {n\over d}\rfloor d) \times (\lceil {w\over q} \rceil+n-\lfloor {n\over d}\rfloor d)$, respectively, and the decoding matrix of G-AGC is size of $L\times L$ and $\big(L+\lceil{L\over d-s}\rceil (n-\lfloor {n\over d}\rfloor d)\big) \times \big(L+\lceil{L\over d-s}\rceil (n-\lfloor {n\over d}\rfloor d)\big)$, respectively. 
	Since the first $\lfloor {n\over d}\rfloor-1$ groups in G-AGC can also use Vandermonde matrix to reduce complexity, so only the last group needs $O((\cdot)^3)$ time to decode. 
	The comparison of decoding complexity is given in Table \ref{table:comp}, where the approximation  is for the situation that $L,d$ are small and fixed, and $n$ is large. It is seen that the complexity of G-AGC and G-CGC can be approximately in the same order $O(n)$. For small parameters, it is verified in Section \ref{sec:amazon} through simulation that, for average running time, the decoding complexity that  G-AGC introduced is trivial compared to the benefits it brings. 
		\begin{table}[!http]
	\center
	\caption{Decoding complexity comparison of G-AGC and G-CGC.}\label{table:comp}
		\begin{tabular}{|c|c|}\hline
			 & 	Decoding complexity \\ \hline
	G-AGC	&	$O((\lfloor {n\over d}\rfloor-1)\cdot L^2)+O((L+\lceil{L\over d-s}\rceil (n-\lfloor {n\over d}\rfloor d))^3) \approx O(n)$		\\ \hline
	G-CGC	&	$O((\lfloor {n\over d}\rfloor-1)\cdot \lceil {w\over q} \rceil^2)+O((\lceil {w\over q} \rceil+n-\lfloor {n\over d}\rfloor d)^2) \approx O(n)$	\\ \hline
		\end{tabular}
   \end{table}


\section{Numerical Comparison}\label{sec:Numerical}
In this section, we compare the performance of CGC, AGC, and their group versions  G-CGC, G-AGC schemes under a model that the system restarts part or all of the stragglers periodically until the master can decode the full gradient, which will be referred to as \emph{restart model}, as did in \cite{AS19}.
\subsection{Restart Model}

The system operates in multiple periods, where each period of $t$ seconds is referred to as an epoch, and the periods will be referred to as epoch 0,1,$\ldots$, sequentially. Each epoch is partitioned into two phases, a computation phase of $t_{\rm{cp}}$ seconds and a communication phase of $t_{\rm{cm}}$ seconds, which satisfies $t_{\rm{cp}}+t_{\rm{cm}}=t$.  That is, by the end of the computation phase, if a worker has accomplished its computation task, it will send coded signals to the master sequentially in the following communication phase, or else, it is seen as a straggler. To simplify the analysis, we assume: 
\begin{enumerate}
\item In each epoch, each restarted worker has a fixed probability $p$ to become a straggler. The event of a worker being a straggler is independent over the workers and epochs.  
\item  The time consumed for communication is proportional to the number of symbols sent to the master, i.e., the communication length. The length $t_{\rm{cm}}$ of a communication phase  is set as the time of transmitting a codeword of length $w$.
\end{enumerate}

The master starts all the $n$ workers in epoch $0$.  At the beginning of the following epoches $i=1,2,\ldots,$ the master restarts part or all of the stragglers otherwise\footnote{For non-group schemes, the master restart all stragglers; for group schemes, only the stragglers in the groups that fail to decode their partial gradients will be restarted.}. At the end of epoch $i$, the master attempts to decode the full gradient. 
If success, it broadcasts a stop signal and updates the parameter $w$, then it goes to the next iteration.
Otherwise, it will launch epoch $i+1$.
Define the  Average Execution Time (AET) of  a scheme to be the expectation of the time for the  master to decode the full gradient. In the following, we will first evaluate the AET of the above mentioned schemes. Notice that, for a $(n,\mu,w)$ system, all the schemes choose $k=n$  and $d=\lfloor n\mu\rfloor$, that is, each worker needs to compute the partial gradients over $d$ data subsets.
 
For convenience of analysis, we first consider a system with $u$ workers, which restarts all stragglers until the number of stragglers does not exceed threshold $s_{\max}$. Let $S_i, i=0,1,\ldots$ be the number of stragglers in epoch $i$\footnote{If there is no restarted workers in epoch $i$, $S_i\triangleq 0$.}. Then  $S_0,S_1,\ldots,$ forms a Markov chain, satisfying $S_0\geq S_1\geq \ldots$, which must also satisfy : $S_{i-1}>s_{\max}$ if $S_i >0$ for all $i>0$. Therefore, the transition probability is
\begin{eqnarray}
&& \hspace{-3mm} \Pr(S_i=a_i\,|\,S_{i-1}=a_{i-1},,\ldots,S_0=a_0) \notag \\ 
&=& \hspace{-3mm} \Pr(S_i=a_i\,|\,S_{i-1}=a_{i-1})\label{Eqn_Markov} 	\\
&=& \hspace{-3mm} \left\{\begin{array}{ll}
\hspace{-2mm} {a_{i\!-\!1}\choose a_i}p^{a_{i}}(1\!-\!p)^{a_{i\!-\!1}-a_i}, \! &\textnormal{if}~a_{i\!-\!1}\!>\!s_{\max}, 0\!\leq \!a_i\! \leq \! a_{i\!-\!1}  \notag \\
\hspace{-2mm} 1, &\textnormal{if}~a_{i\!-\!1}\leq s_{\max}~\textnormal{and}~a_i=0  \notag \\
\hspace{-2mm} 0, \! &\textnormal{else}
\end{array}\right., \label{}
\end{eqnarray}
where $i\geq 1$, and the  the probability distribution of $S_0$ is
\begin{eqnarray*}
\Pr(S_0=a_0)={u\choose a_0}p^{a_0}(1-p)^{u-a_0},\quad a_0\in[0:u].
\end{eqnarray*}
 
Let $Y$ be the first epoch where the number of stragglers does not exceed $s_{\max}$. Define
\begin{IEEEeqnarray*}{c}
h(u,i,s,s_{\max})\triangleq\Pr(Y=i,S_i=s),\quad s\in[0:s_{\max}].
\end{IEEEeqnarray*}
In fact, $h(u,i,s,s_{\max})$ can be evaluated in \eqref{eqn:numerical_agc} at the top of next page, whose proof is given in Appendix. This evaluation is very useful for our following AET analysis.
\begin{figure*}
\begin{normalsize}
	\begin{eqnarray}
h(u,i,s,s_{\max})=
\left\{\begin{array}{ll}
{u\choose s}p^s(1-p)^{u-s}, &\textnormal{if}~{i=0}\\
\sum_{s_{\max}<a_{i-1}\leq \ldots\leq a_0\leq u}\frac{u!}{(u-a_0)!(a_0-a_1)!\ldots(a_{i-1}-s)! s!} \cdot p^{\sum_{j=0}^{i-1}a_j+s}(1-p)^{u-s}, &\textnormal{if}~{i\geq 1}
\end{array}
\right. \label{eqn:numerical_agc}
\end{eqnarray}
\end{normalsize}
\hrulefill
\end{figure*}

\subsection{Evaluation of Average Execution Times}

\subsubsection{The AET of CGC and AGC schemes} 
We use A to denote one of CGC and AGC schemes.
Let $\bm c^\textnormal{A}=(c_0^{\textnormal{A}},c_1^{\textnormal{A}},\ldots,c_{s_{\max}^\textnormal{A}}^{\textnormal{A}})$ be the communication vector achieved by scheme A, where $s_{\max}^{\textnormal{A}}$ is the largest number of stragglers that scheme A can resist, and $ c_{s}^{\textnormal{A}}$ is the communication cost when there are $s$ stragglers. The system will restart all the stragglers in the previous epoch at the beginning of the current one, as long as the full gradient has not been decoded.

 Let $Y^{\textnormal{A}}$ be the epoch in which the master decodes the full gradient, and $S_{i}^{\textnormal{A}}$ be the number of stragglers in the $i$-th epoch. 
Recall that in each epoch where the master cannot decode the gradients, all workers need to wait $t=t_{\textnormal{cp}}+t_{\textnormal{cm}}$ seconds for the master to recognize the stragglers and restart them. If the master decodes at the $i$-th epoch, there are $i$ epochs of restart which takes $i\cdot t$ seconds, and the number of stragglers in last epoch satisfies $s\leq s_{\textnormal{max}}$, which takes  $t_{\textnormal{cp}}+c_s t_{\textnormal{cm}}$ seconds according to the value of $s$. That is, the time for the case of $Y_A=i,S_i^A=s$ is $t_{\textnormal{cp}}+i\cdot t+c_s t_{\textnormal{cm}}$ seconds.

 As a result, the AET of scheme A is 
\begin{IEEEeqnarray}{rCl}
&&\mathbf{E}[T^{\textnormal{A}}]\notag\\
 &=& \sum_{i=0}^\infty\sum_{s=0}^{s_{\max}^{\textnormal{A}}}\Pr(Y^{\textnormal{A}}=i,S^{\textnormal{A}}_i=s)\cdot( t_{\rm{cp}}+c_s^{\textnormal{A}}\cdot t_{\rm{cm}}+i\!\cdot \!t) \notag \\
&=&  t_{\rm{cp}}+ \sum_{i=0}^\infty\sum_{s=0}^{s_{\max}^{\textnormal{A}}}\Pr(Y^{\textnormal{A}}=i,S^{\textnormal{A}}_i=s)\cdot(c_s^{\textnormal{A}} t_{\rm{cm}}+i\!\cdot\! t)\label{eqn:compute:AET}\\
&=& t_{\rm{cp}}+ \sum_{i=0}^\infty\sum_{s=0}^{s_{\max}^{\textnormal{A}}}h(n,i,s,s_{\max}^{\textnormal{A}})\cdot(c_s^{\textnormal{A}} t_{\rm{cm}}+i\cdot t)\label{eqn:AET}
\end{IEEEeqnarray}
Therefore, by plugging the communication vectors \eqref{eqn:c:q} and \eqref{eqn:c:opt} into \eqref{eqn:AET} respectively, we obtain the AET of CGC and AGC as
\begin{IEEEeqnarray*}{rCl}
\mathbf{E}[T^{\textnormal{CGC}}]
&=& t_{\rm{cp}}\!+\! \sum_{i=0}^\infty\sum_{s=0}^{s_{\max}^{\textnormal{CGC}}}h(n,i,s,s_{\max}^{\textnormal{CGC}})\! \cdot\! \Big({1\over d\!-\!s_{\max}^{\textnormal{CGC}}}\!\cdot \!t_{\rm{cm}}+i\!\cdot \!t\Big),\\
\mathbf{E}[T^{\textnormal{AGC}}]
&=&  t_{\rm{cp}}+ \sum_{i=0}^\infty\sum_{s=0}^{d-1} h(n,i,s,d-1)\! \cdot \!\Big({1\over d-s}\cdot t_{\rm{cm}}+i\cdot t\Big)	,	
\end{IEEEeqnarray*}
where $s_{\max}^{\textnormal{CGC}}$ is determined by the design parameter $q$ and satisfies $s_{\max}^{\textnormal{CGC}} \leq d-{w\over q}$ as in \eqref{eqn:bound:2}.
When $s_{\max}^{\textnormal{CGC}}=d-1$, it is obvious that $\mathbf{E}[T^{\textnormal{AGC}}] < \mathbf{E}[T^{\textnormal{CGC}}]$, and our extensive numerical results indicate that it also holds for $s_{\max}^{\textnormal{CGC}}<d-1$.

\subsubsection{The AET of G-CGC and G-AGC schemes}

Following the above notations, we use G-A to denote the group version of scheme A, where  A is one of CGC or AGC schemes.
Recall that in Section \ref{sec:Group}, the workers are partitioned into $m+1$ groups $\mathcal{I}_0,\mathcal{I}_1,\ldots,\mathcal{I}_m$, where $m=\left\lfloor \frac{n}{d}\right\rfloor -1$. Each of the first $m$ groups $\mathcal{I}_0,\ldots,\mathcal{I}_{m-1}$ contains $d$ workers, and forms  a $(d,1,w)$ subsystem. The last group $\mathcal{I}_m$ forms a $(n-dm, \frac{d}{n-dm},w)$ subsystem.  
Notice that, by plugging the above subsystem parameters into \eqref{eqn:c:q} or \eqref{eqn:c:opt},
all the subsystems using scheme A (CGC or AGC)
achieve the same communication vector $\bm c^{\textnormal{A}}=(c_0^{\textnormal{A}},c_1^\textnormal{A},\ldots,c_{s_{\max}}^\textnormal{A})$. The master can decode the full gradient if and only if all the partial gradients $\overline{g}^{(0)},\overline{g}^{(1)},\ldots,\overline{g}^{(m)}$ computed by the subsystems can be decoded. Therefore, for each subsystem $j$, the master will restart all the stragglers in each epoch until that the number of stragglers in subsystem $j$ does not exceed $s_{\max}^{\textnormal{A}}$.

Let $Y_j^{\textnormal{G-A}}$ be the epoch in which, the $j$-th subsystem can decode the partial gradient $\overline{g}^{(j)}$, and $S_{j,i}^{\textnormal{G-A}}$ be the number of the stragglers of the $j$-th subsystem in  epoch $i$. Thus, the master  decodes the $j$-th partial gradient $\overline{g}^{(j)}$ in epoch $i$ if and only if $Y_j^{\textnormal{G-A}}=i$ and $S_{j,i}^{\textnormal{G-A}}\leq s_{\max}^{\textnormal{A}}$.
Let $Y^{\textnormal{G-A}}=\max\{Y_0^{\textnormal{G-A}},Y_1^{\textnormal{G-A}},\ldots,Y_m^{\textnormal{G-A}}\}$ be the epoch at which the master decodes the full gradient, and $S^{\textnormal{G-A}}_{i}=\max\{S_{0,i}^{\textnormal{G-A}},S_{1,i}^{\textnormal{G-A}},\ldots,S_{m,i}^{\textnormal{G-A}}\}$ be the maximum number of stragglers over the subsystems in epoch $i$. Then the master decodes at epoch $i$  if and only if $Y^{\textnormal{G-A}}=i$ and $S_i^{\textnormal{G-A}}\leq s_{\max}^{\textnormal{A}}$. Following the same steps as in \eqref{eqn:compute:AET}, the AET of G-A scheme is:
\begin{IEEEeqnarray}{rCl}
\mathbf{E}[T^{\textnormal{G-A}}] =
\quad t_{\rm{cp}}\!+\!\sum_{i=0}^\infty\sum_{s=0}^{s_{\max}^{\textnormal{A}}}\Pr(Y^{\textnormal{G-A}}=i,S^{\textnormal{G-A}}_i=s)\!\cdot\!(c_s^{\textnormal{A}}\cdot t_{\rm{cm}}\!+\!i\cdot t). \label{eqn:AET_Group}
\end{IEEEeqnarray}

In fact, for each $i\geq 0$ and $s\in[0:s_{\max}^{\textnormal{A}}]$, we can evaluate $\Pr(Y^{\textnormal{G-A}}=i,S^{\textnormal{G-A}}_i=s)$ in \eqref{eqn:eva} at the top of next page, and its derivation is deferred to Appendix.
\begin{figure*}[ht]
\vspace{-6mm}

\begin{IEEEeqnarray}{rCl}
&&\Pr(Y^{\textnormal{G-A}}=i,S^{\textnormal{G-A}}_i=s) \notag\\
 &=&\left\{\begin{array}{ll}
\Big(\sum_{b=0}^{i-1}\sum_{a=0}^{s_{\max}^{\textnormal{A}}} h(d,b,a,s_{\max}^{\textnormal{A}})+h(d,i,0,s_{\max}^{\textnormal{A}})\Big)^m \\
\quad \cdot \Big(\sum_{b=0}^{i-1}\sum_{a=0}^{s_{\max}^{\textnormal{A}}} h(n-md,b,a,s_{\max}^{\textnormal{A}})+h(n-md,i,0,s_{\max}^{\textnormal{A}})\Big)-\\
\quad \Big(\sum_{b=0}^{i-1}\sum_{a=0}^{s_{\max}^{\textnormal{A}}} h(d,b,a,s_{\max}^{\textnormal{A}})\Big)^m\Big(\sum_{b=0}^{i-1}\sum_{a=0}^{s_{\max}^{\textnormal{A}}} h(n-md,b,a,s_{\max}^{\textnormal{A}})\Big),&\textnormal{if}~ s=0 \\
\bigg(\sum_{b=0}^{i-1}\sum_{a=0}^{s_{\max}^{\textnormal{A}}}h(d,b,a,s_{\max}^{\textnormal{A}})+\sum_{a=0}^{s-1}h(d,i,a,s_{\max}^{\textnormal{A}})\bigg)^m \\
\quad\cdot \bigg(\sum_{b=0}^{i-1}\sum_{a=0}^{s_{\max}^{\textnormal{A}}}h(n-md,b,a,s_{\max}^{\textnormal{A}})+\sum_{a=0}^{s-1}h(n-md,i,a,s_{\max}^{\textnormal{A}})\bigg),
&\textnormal{if}~1\leq s\leq s_{\max}^{\textnormal{A}}
\end{array}
\right.\label{eqn:eva}
\end{IEEEeqnarray}
\hrulefill
\end{figure*}
Therefore, 
\begin{IEEEeqnarray*}{rCl}
&&\mathbf{E}[T^{\textnormal{G-AGC}}] = \quad t_{\rm{cp}}\!+\! \sum_{i=0}^\infty\sum_{s=0}^{d-1}\Pr(Y^{\textnormal{G-AGC}}\!=\!i,S^{\textnormal{G-AGC}}_i\!=\!s) \Big({t_{\rm{cm}}\over d\!-\!s}\!+\!i\!\cdot\! t\Big),	 	\\
&&\mathbf{E}[T^{\textnormal{G-CGC}}]\! = \!t_{\rm{cp}}\!+ \!\sum_{i=0}^\infty \! \sum_{s=0}^{s_{\max}^{\textnormal{CGC}}}\!\! \Pr(Y^{\textnormal{G-CGC}}\!\!\!=\!i,S^{\textnormal{G-CGC}}_i\!\!\!=\!s)\Big(\!{t_{\rm{cm}}\over d\!-\!s_{\max}^{\textnormal{CGC}}}\!+\!i\!\cdot \!t\Big),	
\end{IEEEeqnarray*}
where $\Pr(Y^{\textnormal{G-AGC}}=i,S^{\textnormal{G-AGC}}_i=s)$ and $\Pr(Y^{\textnormal{G-CGC}}=i,S^{\textnormal{G-CGC}}_i=s)$ are evaluated according to \eqref{eqn:eva}. 
Apparently, $\mathbf{E}[T^{\textnormal{G-AGC}}] < \mathbf{E}[T^{\textnormal{G-CGC}}]$ when $s_{\max}^{\textnormal{CGC}}=d-1$, and extensive numerical results indicate that it also holds for other cases.

\subsection{Numerical  Comparison}
\begin{figure}[!htb]
\setlength{\abovecaptionskip}{-0.1cm} 
	\centering
	\includegraphics[width=0.5\textwidth]{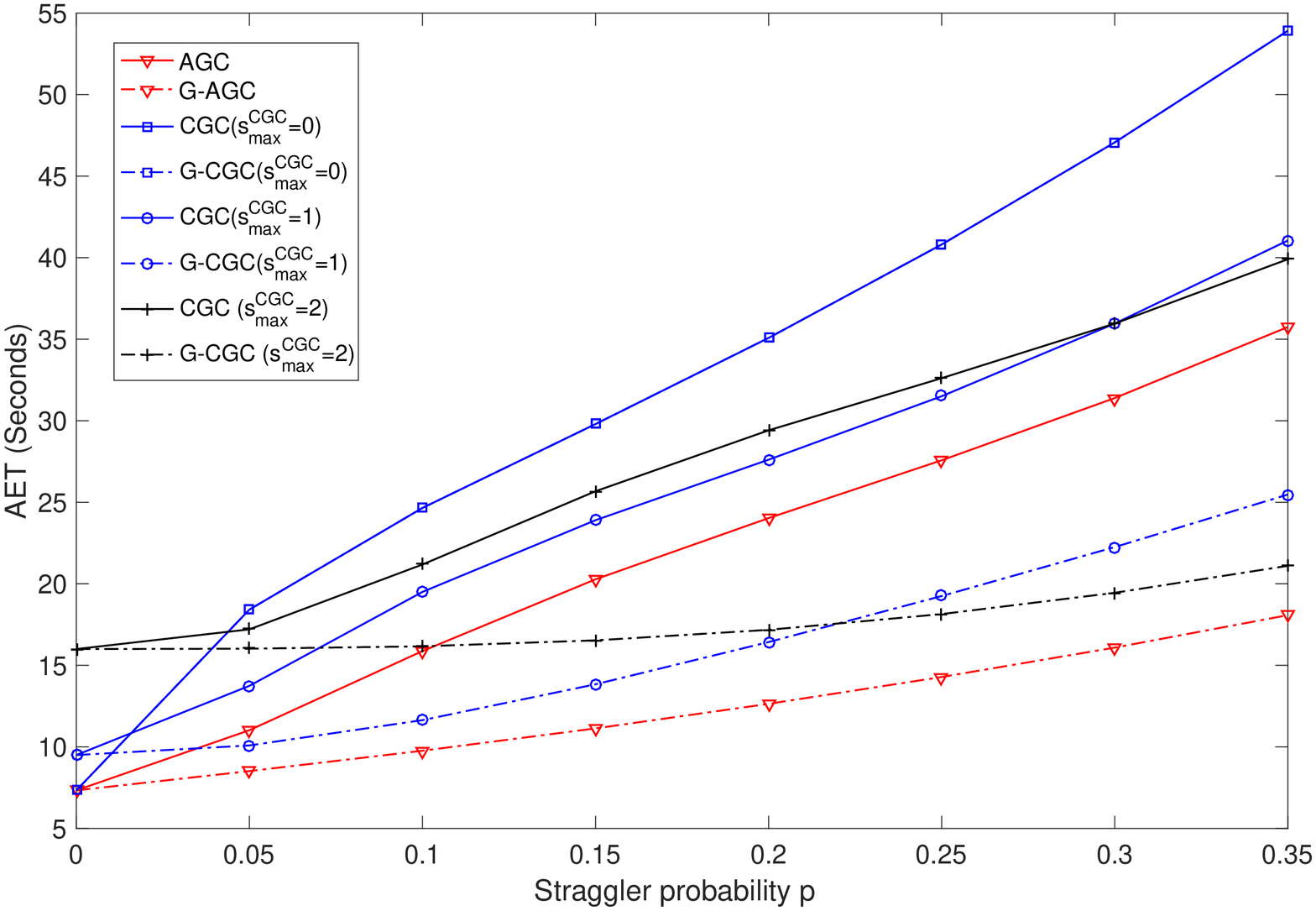}
	\caption{Average Execution Time with $n=20,d=3,t_{\rm{cm}}=13,t_{\rm{cp}} = 3$ and $t = 16$.}\label{fig:runtime_matlab}
\end{figure}

In Fig. \ref{fig:runtime_matlab}, we plot the AET of the schemes given above as functions of straggler probability $p$ for a system with $n=20$ workers, and $\mu=\frac{3}{20}$, i.e., each worker stores $d=\lfloor n\mu \rfloor =3$  of the $20$ data subsets, where  $t_{\rm{cm}}= 13$ and $t_{\rm{cp}} = 3$. For CGC and G-CGC schemes, for each  $s_{\max}^{\textnormal{CGC}}\in\{0,1,2\}$, we choose the smallest $q$ satisfying \eqref{eqn:bound:2} to form their communication vector. 

From Fig. \ref{fig:runtime_matlab}, we have the following observations. 
\begin{enumerate}
\item  For each of CGC with $s_{\max}^{\textnormal{CGC}}\in\{1,2\}$ and AGC scheme, the group version of these schemes always outperforms the original ones, this verifies the fact that the group approach helps to improve the robustness. Note that when $s_{\max}^{\textnormal{CGC}}=0$, G-CGC has the same performance as CGC since they both can not tolerate any straggler.

\item Comparing the three CGC (or G-CGC) schemes (i.e., $s_{\max}^{\textnormal{CGC}}\in\{0,1,2\}$), when $p$ is
\begin{itemize}
\item  extremely small (approximately $p< 0.01$), which makes it close to straggler-free system;
\item moderate  (approximately $0.015<p<0.3$ for CGC,  $0.01<p<0.23$ for G-CGC);
\item large  ($p>0.3$ for CGC, $p>0.23$ for G-CGC),  where the number of stragglers is typically larger than $1$,
\end{itemize}
the best scheme is the one with $s_{\max}^{\textnormal{CGC}}$ being $0,1,2$ respectively. 
The difference of their performances originates from the fact that CGC scheme is designated to effectively reduce the communication cost for a fixed number of stragglers determined by $s_{\max}^{\textnormal{CGC}}$. When the real-time number of stragglers does not match its parameter, the performance is deteriorated. 
\item Comparing AGC (resp. G-AGC) with CGC (resp. G-CGC) schemes, AGC (resp. G-AGC) outperforms all CGC (resp. G-CGC) in all regimes. This is because unlike CGC (resp. G-CGC),  AGC (resp. G-AGC) adaptively reduces the communication cost according to the real-time number of stragglers. 
  Thus, given that the number of stragglers in the system is random, the AET of AGC (resp. G-AGC) has advantage over CGC (resp. G-CGC) schemes. 
\end{enumerate}

As a result, G-AGC scheme simultaneously enjoys the benefits of adaptively communication cost reduction according to the real-time number of stragglers and robustness provided by group approach, which performs the best for all $p$.

\section{Implementation on Amazon EC2 clusters}\label{sec:amazon}

In this section,  we conduct an image recognition training in Amazon EC2 cluster, where we use Pytorch with mpi4py package on m4.2xlarge instances for implementation.
As we observe from the experiments, the numerical stability deteriorates very quickly as the number of workers $n$ becomes larger than 20 such that the training can not even converge if it is larger than $24$. Therefore, we choose $n=20$. 
The settings of the training model summarized in Table.\ref{table:data} follow the one in \cite{ChenWCP18} except for the size of cluster.
Specifically, we employ ResNet-18 model to train the Cifar-10 dataset with training set of 50000 data points, and test set of the rest 10000 ones. The parameter $w$ in the experiment is of size 11173962. In addition, a mini-batch Stochastic Gradient Descent (SGD) method is adopted, whose training batch size in each iteration is set to be 180 and learning rate is fixed as 0.1. 
For all the schemes, each training batch are split into $k=20$ data subsets, each containing ${180\over 20}=9$ data points. The parameter $\mu=\frac{d}{n}=0.15$, i.e., each worker can store up to $d=3$ data subsets for each training batch.
In each round of communication, we use MPI.Isend() function to send signals and MPI.Irecv() function to receive them, where the functions automatically use TCP/IP to transmit and receive data.


  \begin{table}[!http]
	\center
	\caption{Settings of the training.}\label{table:data}
		\begin{tabular}{|c|c|c|}\hline
	Dataset	 & Cifar-10 \\ \hline
	Size of cluster	& 20  \\ \hline
	Data points	& 60,000		\\ \hline
	Parameter $w$	& 11,173,962	\\ \hline
	Model	& ResNet-18	  \\ \hline
	Optimizer	& SGD\\ \hline
	Learning rate	& 0.1  \\ \hline
	Batch size	& 180  \\ \hline
		\end{tabular}
   \end{table}

We will compare the following schemes:
\begin{enumerate}
	\item The uncoded scheme, where the data subsets are evenly distributed to all workers and the master waits to combine the results from all of them. Notice that for uncoded scheme, each worker only needs to compute one data subset per iteration.
	\item The CGC scheme \cite{YeA18} and its group version G-CGC with $s_{\max}^{\textnormal{CGC}}\in\{0,1,2\}$, where all of them will compute $d=3$ data subsets per iteration, and the communication length $q$ in each case is determined by the least $q$ satisfying (9). That is, $q = {w\over 3}$ for $s_{\max}^{\textnormal{CGC}}=0$, and $q = {w\over 2}$ for $s_{\max}^{\textnormal{CGC}}=1$, and $q=w$ for $s_{\max}^{\textnormal{CGC}}=2$.
	\item The AGC scheme and its group version G-AGC. Similarly, both of them needs to compute $d=3$ data subsets per iteration. Then, we choose  $L= \mathrm{lcm}(1,2,3)=6$ for AGC and G-AGC, i.e., each gradient vector is split into $L=6$ sub-vectors of length ${w\over 6}$. Therefore, each worker will send at most 6 rounds of signals by calling MPI.Isend() function at most $6$ times until the master broadcasts the stop signal. 
\end{enumerate}

\begin{figure}[!htb]
	\setlength{\abovecaptionskip}{-0.1cm}
	\centering
		\includegraphics[width=0.4\textwidth]{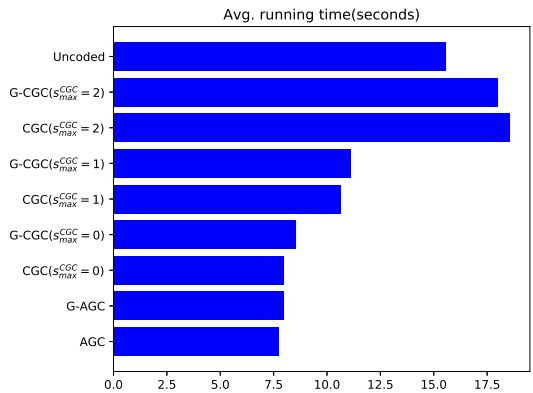}
	\caption{Average running time in normal Amazon environment with $n=20$ workers.}
	\label{fig:avg_n20_a}
\end{figure}

All the schemes were trained for 2000 iterations so that they can achieve the same designated testing accuracy $80\%$. 
Firstly, we implemented the schemes in normal Amazon environment, where the stragglers are only caused by EC2 clusters. The average running time of all schemes is shown in Fig. \ref{fig:avg_n20_a}. During our experiments, m4.2xlarge instances are very stable such that no straggling worker is observed. As a result, it can be seen that for AGC or any particular CGC scheme, its group version has almost the same running time as itself since the advantage of robustness to stragglers of the group scheme is not evident. For CGC and G-CGC schemes, the smaller $s_{\max}^{\textnormal{CGC}}$, the better performance, due to the communication savings. As expected, CGC with $s_{\max}^{\textnormal{CGC}}=2$ performs worse than uncoded scheme, since it sends the same size of vector but introduces computation redundancy, while the other coded schemes can have the size reduction to decelerate communication time. 
AGC and G-AGC perform as good as CGC with $s_{\max}^{\textnormal{CGC}}=0$, since they have the same lowest communication cost. 

To observe the effect of stragglers, we fulfilled another two experiments, in which, each worker has probability $p=0.05$ and $p=0.1$ to become a straggler to simulate the situation where the number of stragglers in average is approximately $s_{\max}^{\textnormal{A}}$. 
Particularly, to simulate the straggler behavior, each worker generates a random variable uniform in $[0:1]$, and compare it with the given probability $p$. If the variable is smaller than $p$, the worker becomes a straggler and keeps idle, otherwise, it will finish the task and wait for next iteration. 
In such systems, the full gradient is computed by the restart model described in Section VII, i.e.,  after a period of time $t$, the master will restart all stragglers for non-group schemes, or all the stragglers in the groups that has not computed their partial gradients for group schemes. 

\begin{figure}[!htb]
	\setlength{\abovecaptionskip}{-0.1cm}
	\centering
		\subfigure[The straggler probability $p=0.05$.]
		{ \label{fig:avg_n20_b}
		\includegraphics[width=0.4\textwidth]{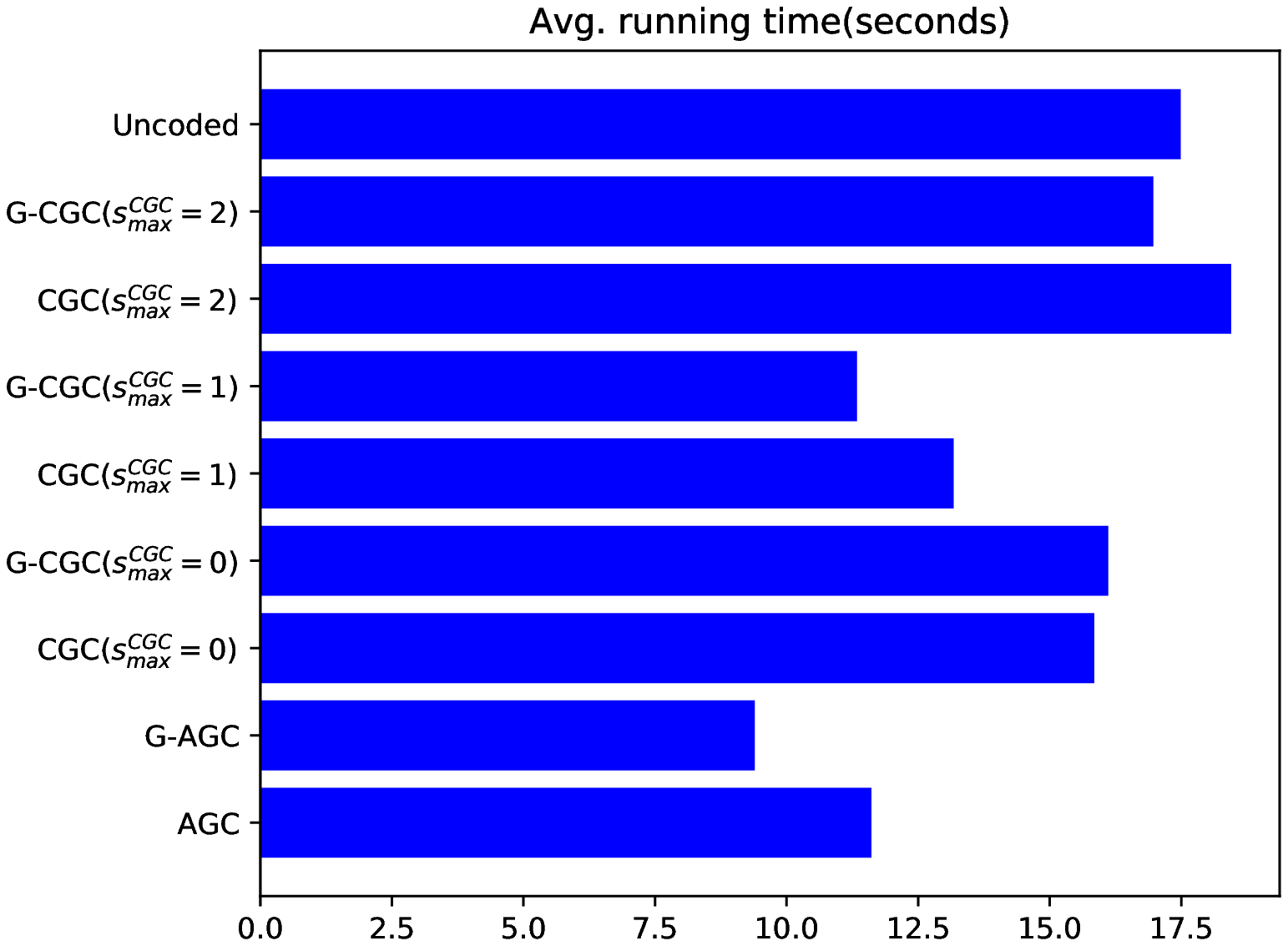}
		}
		\subfigure[The straggler probability $p=0.1$.]
		{ \label{fig:avg_n20_c}
		\includegraphics[width=0.4\textwidth]{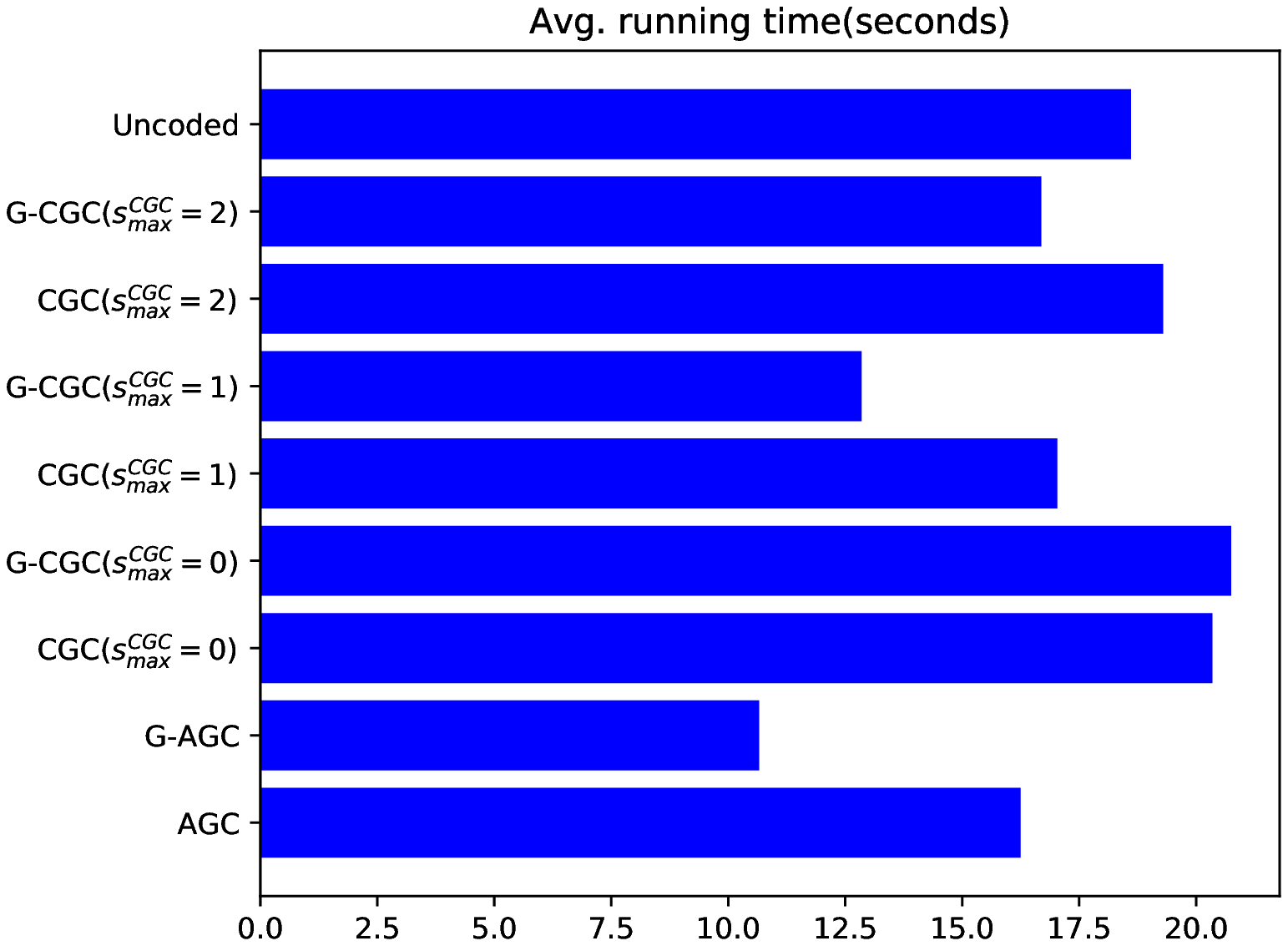}
		}
	\caption{Average running time per iteration with $n = 20$ workers.}
	\label{fig:run_time_n20}
\end{figure}

To determine $t$ for the restart model, we need to clarify the time duration of an epoch. In fact, each iteration can be divided into six phases: broadcast $w$ (refer to Bcast), calculate the gradient (refer to Comp), send the gradient (refer to Comm), encode (refer to Enc), decode (refer to Dec), and update $w$ (refer to Update). A  
detailed time duration of some typical iterations is listed in Table. \ref{table:time_compare_two_epochs} and Table. \ref{table:time_compare_AGC}, where we only list the non-group schemes since the ones in group versions are similar.
It is seen that the communication time  dominates  epoch 0 but soon dramatically decreases from epoch 1.  According to our test, this phenomenon is caused by a reduced network congestion, i.e., there are only few stragglers restarted from epoch 1 and hence the transmission  is speeded up within a fixed bandwidth.
In fact, a similar network congestion phenomenon was observed in \cite{TandonLDK17} as well, where the system was bottlenecked by the number of incoming connections, i.e., all workers were trying to talk to the master concurrently. Thus, for epoch $i$ in our training, we set $t=16$ if $i=0$ and $t=4$ otherwise, according to the total time of each epoch in tables \ref{table:time_compare_two_epochs} and \ref{table:time_compare_AGC}. 
Then, the average running time of all schemes for $p=0.05$ and $p=0.1$ are shown in Fig. \ref{fig:run_time_n20}.

   \begin{table*}[!http]
	\centering
	\small
	\caption{The time duration of different phases in uncoded and CGC schemes.}\label{table:time_compare_two_epochs}
		\begin{tabular}{|c|c|c|c|c|c|c|c|c|c|c|c|c|}
		\hline
		&  \multicolumn{1}{c|}{\multirow{2}{*}{Bcast}} & \multicolumn{4}{|c|}{Epoch $i=0$} & \multicolumn{4}{|c|}{Epoch $i>0$} & \multicolumn{1}{c|}{\multirow{2}{*}{Dec}} &  \multicolumn{1}{c|}{\multirow{2}{*}{Update}}	\\ \cline{3-10}
		&   &	Comp	&	Comm  &	Enc  &Total & Comp	&	Comm 	& Enc  &Total &  &  \\ \hline
		Uncoded & 1.15 & 0.82 &	12.59  &	 0.15 	&  13.56  &	0.82 & 1.08 & 0.14 & 2.04 & 0.47 & 0.04  \\	\hline 
		CGC($s_{\max}^{\textnormal{CGC}}=0$) & 1.15 & 2.45  & 3.17 & 0.33	 &	5.95 & 2.45 & 	0.26 &	0.32 & 3.03 & 0.79 & 0.04  \\	\hline
		CGC($s_{\max}^{\textnormal{CGC}}=1$) & 1.15 & 2.45 &	 5.95  & 0.43  & 8.83 & 2.45 & 	0.57 & 0.43	 & 3.45 & 0.95 & 0.04  \\	\hline
		CGC($s_{\max}^{\textnormal{CGC}}=2$) & 1.15 & 2.45 & 12.56  & 0.44  & 15.45 &	2.45 & 1.12	& 0.43  & 4 & 1.58 & 0.04 	\\	\hline
		\end{tabular}
   \end{table*} 
   
   \begin{table*}[!http]
	\centering
	\small
	\caption{The time duration of different phases in AGC scheme.}\label{table:time_compare_AGC}
		\begin{tabular}{|c|c|c|c|c|c|c|c|c|c|c|c|c|c|}
		\hline
	& \multicolumn{1}{c|}{\multirow{2}{*}{Number of stragglers}}	& \multicolumn{1}{c|}{\multirow{2}{*}{Bcast}} & \multicolumn{4}{|c|}{Epoch $i=0$} & \multicolumn{4}{|c|}{Epoch $i>0$} &  \multicolumn{1}{c|}{\multirow{2}{*}{Dec}} & \multicolumn{1}{c|}{\multirow{2}{*}{Update}} \\ \cline{4-11}
	&	 &  &	Comp	&	Comm  &	Enc  &Total & Comp	&	Comm 	& Enc  &Total &  & \\ \hline
\multirow{4}{*}{AGC}	&	$s=0$ & 1.15 & 2.45 & 3.12 & 0.37  & 5.94 & -  &	- &	-  & - & 0.63 & 0.04  \\	\cline{2-13}
	&	$s=1$ & 1.15 & 2.45 & 6.01 & 0.47  & 8.93 & -  &	- &	-  & - & 0.87 & 0.04  \\		\cline{2-13}
	&	$s=2$ & 1.15 & 2.45 & 12.51 & 0.92  & 15.88 & -  &	- &	-  & - & 1.67 & 0.04  \\		\cline{2-13}
	&	$s>2$ & 1.15 & 2.45 & 12.47 & 0.94  & 15.86 & 2.45  &	0.27 &	0.32  & 3.04 & 0.65 & 0.04 \\		\hline
		\end{tabular}
   \end{table*}

To further verify the  experiment results, we made a numerical calculation.  We slightly modified \eqref{eqn:AET} and \eqref{eqn:AET_Group} to fit the parameters in our experiments as follows 
	\begin{eqnarray}\label{eqn:numerical_new}
		\mathbf{E}[T^{\textnormal{A}}] = t_{\rm{cp}} \!+\! t_{\rm{ov}} +  \sum_{i=0}^\infty\sum_{s=0}^{s_{\max}^{\textnormal{A}}}h(n,i,s,s_{\max}^{\textnormal{A}})\!\cdot\!(c_s^{\textnormal{A}} t_{\rm{cm}}^{(i)}\!+\!t^{(\rm{A},\rm{ed})}_s\!+\! i\!\cdot \!t_i),
	\end{eqnarray}
	\begin{eqnarray}\label{eqn:numerical_new_group}
		&&\hspace{-12mm} \mathbf{E}[T^{\textnormal{G-A}}]= t_{\rm{cp}}\!+\! t_{\rm{ov}}+  \sum_{i=0}^\infty\sum_{s=0}^{s_{\max}^{\textnormal{A}}}\Pr(Y^{\textnormal{G-A}}=i,S^{\textnormal{G-A}}_i=s)\!\cdot\!(c_s^{\textnormal{A}} t_{\rm{cm}}^{(i)}\!+\!t^{(\rm{A},\rm{ed})}_s\!+\!i\!\cdot\! t_i),
	\end{eqnarray}
	where $t_{\rm{cm}}^{(i)}$ is the communication time in epoch $i$, and $t_i=16,t_{\rm{cm}}^{(i)}=13$ if $i=0$ and $t_i=4,t_{\rm{cm}}^{(i)}=1$ otherwise, $t_{ov}$ and $t^{(\rm{A},\rm{ed})}_s$ are respectively the time of Bcast and Update and the time of Enc and Dec.  Recall that the group schemes and the corresponding non-group one
have almost the same time duration of different phases. Then, in \eqref{eqn:numerical_new} and \eqref{eqn:numerical_new_group} we use the same $t_{ov}$ and $t^{(\rm{A},\rm{ed})}_s$ given in Tables \ref{table:time_compare_two_epochs} and  \ref{table:time_compare_AGC}.
	Based on \eqref{eqn:numerical_new} and \eqref{eqn:numerical_new_group}, the average running time of each scheme calculated by Matlab is given in Table. \ref{table:numerical}, where it can be seen that the experiment results are indeed consistent with the numerical ones.


   \begin{table}[!http]
	\centering
	\caption{Average running time calculated by Matlab, where $t_{cp}=2.45$, $t_i=16,t_{cm}^{(i)}=13$ is set for epoch $i=0$, and $t_i=4,t_{cm}^{(i)}=1$ is set for $i>0$.}\label{table:numerical}
		\begin{tabular}{|c|c|c|c|c|c|c|c|c|c|c|c|c|}
		\hline
		 &  $p=0.05$ & $p=0.1$ 	\\ \hline
		Uncoded  &	18.39  &	 19.95  \\	\hline
		CGC($s_{\max}^{\textnormal{CGC}}=0$)   &	17.00  & 20.45	  \\	\hline
		G-CGC($s_{\max}^{\textnormal{CGC}}=0$)  &	17.00  & 20.45	  \\	\hline
		CGC($s_{\max}^{\textnormal{CGC}}=1$)  & 14.18  & 17.68  \\	\hline
		G-CGC($s_{\max}^{\textnormal{CGC}}=1$)  & 11.90  & 12.86  \\	\hline
		CGC($s_{\max}^{\textnormal{CGC}}=2$)  & 18.96  & 19.96  	\\	\hline
		G-CGC($s_{\max}^{\textnormal{CGC}}=2$) & 18.67  & 18.70  	\\	\hline
		AGC  & 12.19 & 15.85   \\	\hline
		G-AGC & 9.95 & 11.10   \\	\hline
		\end{tabular}
   \end{table} 
	
   
	
Surprisingly, it is seen from both the experiment results and the numerical ones
 that CGC with $s_{\max}^{\textnormal{CGC}}=2$ performs even worse than uncoded one when $p>0$. This is because the probability that the number of stragglers $s=0$ or $s>2$ is about $43.4\%$ when $p=0.05$ and $44.5\%$ when $p=0.1$, where CGC with $s_{\max}^{\textnormal{CGC}}=2$ has a longer running time  because of the computation redundancy such that the saved time for $1\leq s\leq 2$ can not offset it. 
 Similarly, CGC with $s_{\max}^{\textnormal{CGC}}=0$ and its group version perform worse than uncoded one, since it can not tolerate any straggler and the saved communication time is smaller than the cost of computation redundancy in the scenario of $p=0.1$. 

We also note that the one with $s_{\max}^{\textnormal{CGC}}=1$ performs best in CGCs, since it can tolerate 1 straggler and save some communication cost at the same time. This is to say, both of the communication saving and the straggler tolerant ability are important, which also fits to our experiments that AGC and G-AGC can always perform the best since it adaptively saves the communication cost as much as possible for different number of stragglers.

	It is worthy to mention that, there is a slight difference between the above one and that in Fig. \ref{fig:runtime_matlab}, i.e., CGC with $s_{\max}^{\textnormal{CGC}}=0$ performs better than CGC with $s_{\max}^{\textnormal{CGC}}=2$ when $p=0.05$.
	This is because in theoretical analysis in Section VII, we always assume that the same $t$ and $t_{cm}$ are needed for the restart of all epochs, while in the implementations here, both of the length of epoch and $t_{cm}^{(i)}$ with $i>0$ are much smaller. Therefore, the saved communication time in CGC with $s_{\max}^{\textnormal{CGC}}=0$ can offset the small delay penalty caused by restart.

\begin{figure}[!htb]
	\setlength{\abovecaptionskip}{-0.1cm}
	\centering
		\subfigure[Normal Amazon environment.]
		{ \label{fig:test_ac_n20_a}
		\includegraphics[width=0.3\textwidth]{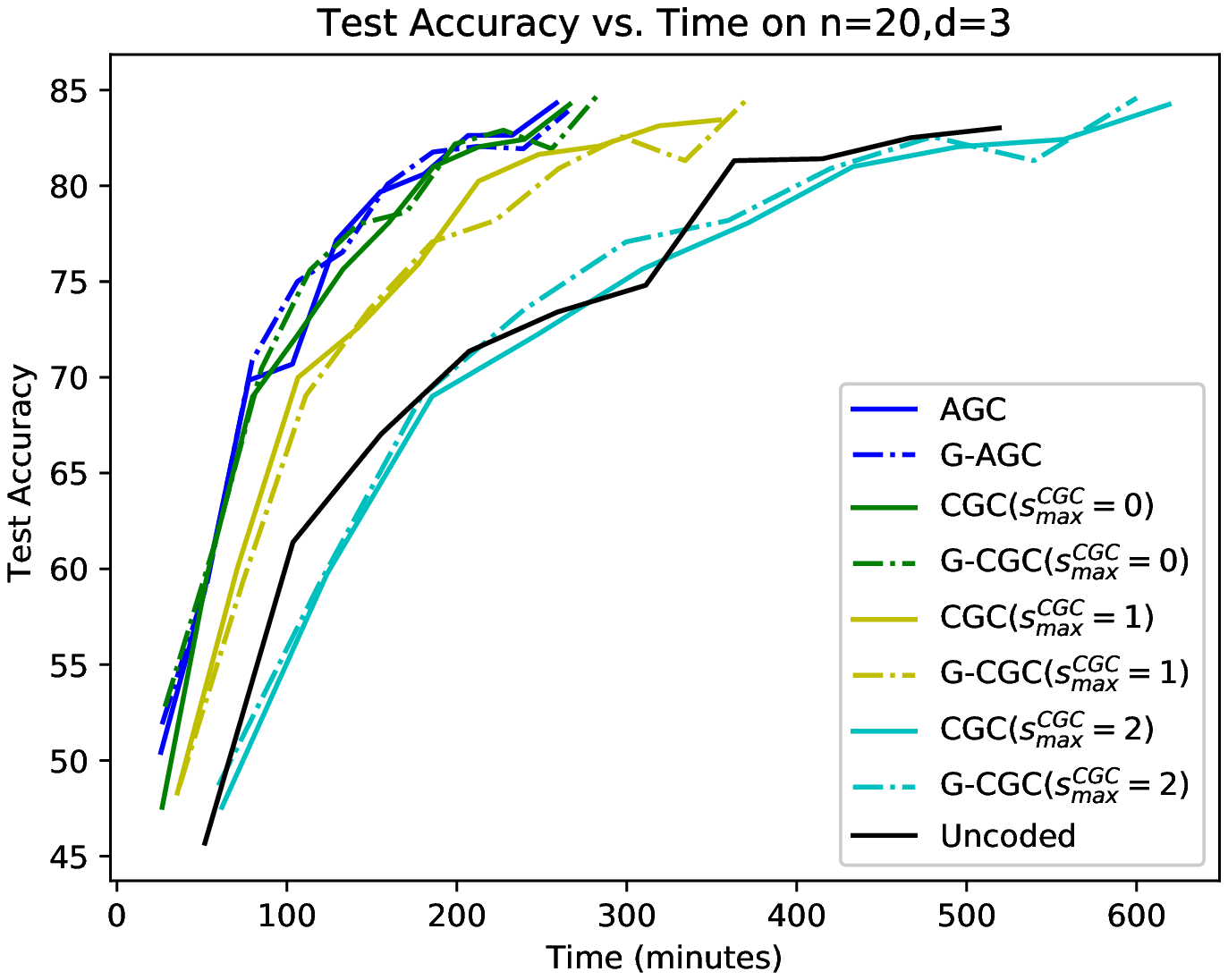}
		}
		\subfigure[The straggler probability $p=0.05$.]
		{ \label{fig:test_ac_n20_b}
		\includegraphics[width=0.3\textwidth]{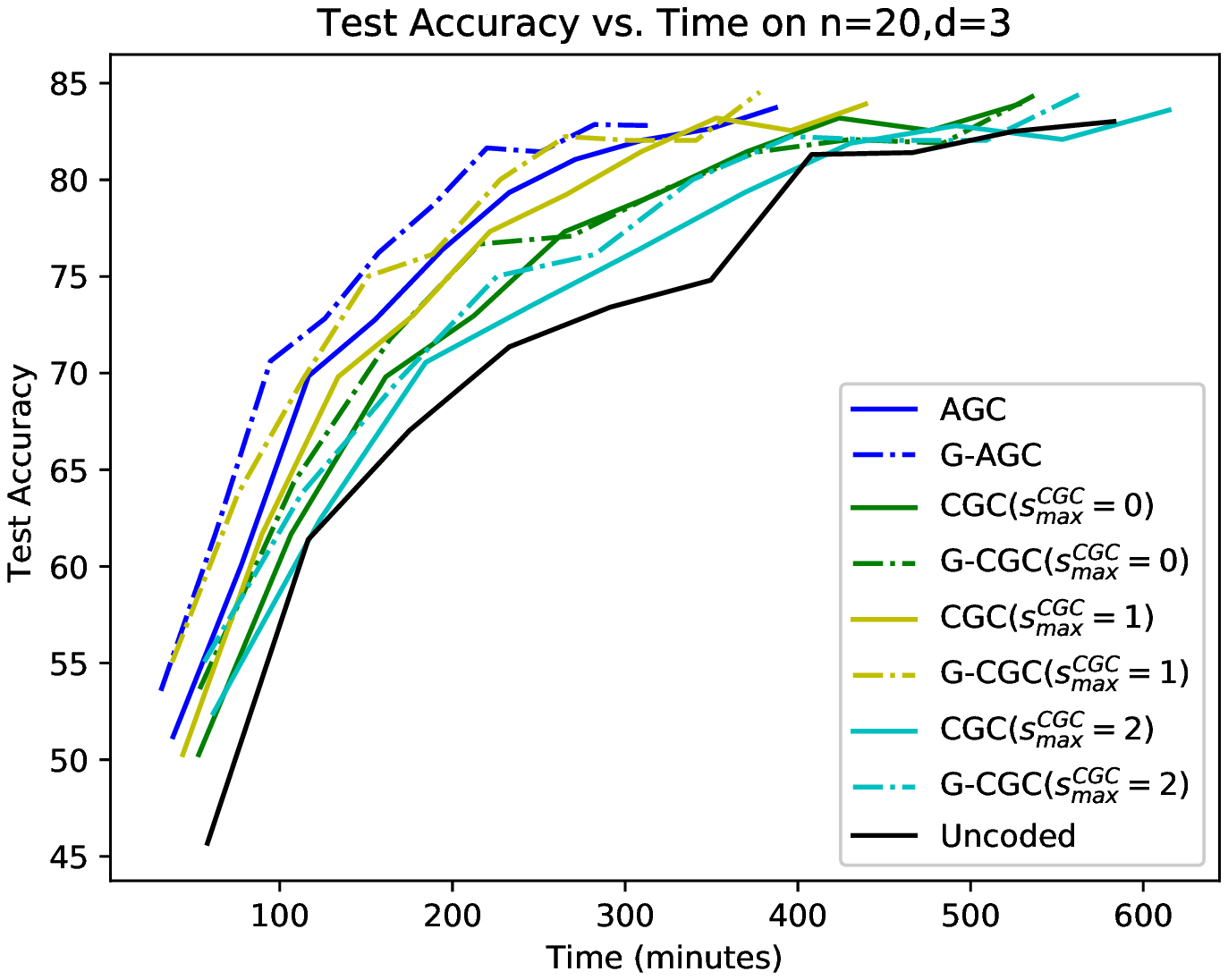}
		}
		\subfigure[The straggler probability $p=0.1$.]
		{ \label{fig:test_ac_n20_c}
		\includegraphics[width=0.3\textwidth]{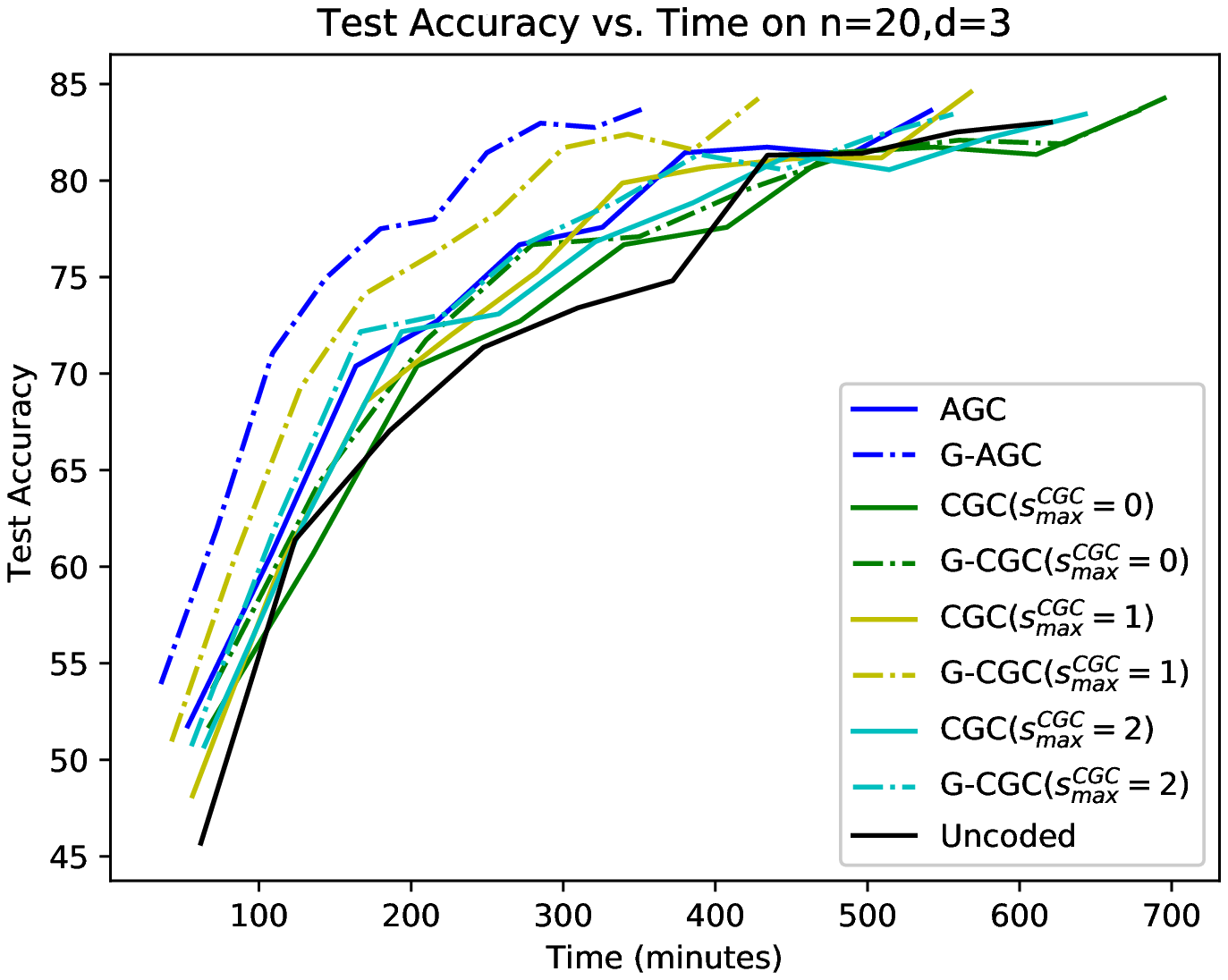}
		}
	\caption{Test accuracy v.s. running time with $n = 20$ workers.}
	\label{fig:test_ac_n20}
\end{figure}

	We also plot the comparison of the test accuracy v.s. running time of all schemes in Fig. \ref{fig:test_ac_n20}.
	The comparison results of different schemes are consistent with the previous figures. In particular, the line for G-AGC is always the leftmost, i.e., it can complete the task with a target accuracy with the shortest time.  

	The above observations verify that CGC or G-CGC scheme needs to choose suitable parameter $s_{\max}^{\textnormal{CGC}}$ to attain
	good performance. In contrast, AGC and G-AGC can achieve better performance in various implementation environments since they can adaptively save communication cost according to the (unknown) real-time number of stragglers.
	
	As a final remark, although AGC can not be used in large cluster due to numerical issues, the group version G-AGC is applicable with better numerical stability. This is because AGC is independently used in its different groups, each having at most $2d-1$ workers, where $d$ is usually small due to the cost of computation redundancy. As we tested, when $d=3$ is chosen, the size of cluster can be more than 40 since the largest group has only 4 workers. Therefore, for a large cluster, we can use G-AGC to avoid the numerical issues and obtain better performance as well.

\section{Conclusion}
In this paper, we investigated the problem of computing gradient through distributed systems.  The proposed AGC scheme achieve the optimal communication cost for any tolerable number of stragglers. Moreover, we incorporated the group idea in \cite{Kadhe20} into AGC to form G-AGC scheme to reduce the complexity and strengthen its robustness to stragglers. The numerical and implementation results demonstrate that our adaptive schemes can achieve the smallest average running time. 
Thus, it is suitable for the practical applications when the real-time number of stragglers is unclear and varying over iterations.

\section*{Acknowledgements}
The authors would like to thank the Associate Editor Prof. Vaneet Aggarwal and the anonymous referees for their constructive suggestions. We are also very grateful to Dr. Min Ye for providing the CGC source code.

\bibliographystyle{IEEEtran}
\bibliography{reference}


\begin{appendix}

\textbf{\textit{Proof of Theorem \ref{thm:numerical}:}}

Given $n,d,L$, when there are $s$ stragglers, we have $r_s=\lceil{L\over d-s}\rceil$ and the encoding matrix $E_{\mathcal{H},\mathcal{H}}^{(\mathcal{F})}$ of the following form
  	\begin{eqnarray}\label{eqn:E}
	E_{\mathcal{H},\mathcal{H}}^{(\mathcal{F})}=\left[
		\begin{array}{ccccccccc}
			E^{(0)}_{\mathcal{F},:} \quad  \bm{0}_{(n-s)\times (n-d)(r_s-1)}	\\
			E^{(1)}_{\mathcal{F},:} \quad \bm{0}_{(n-s)\times (n-d)(r_s-2)} 		\\
			\vdots	\\
			E^{(r_s-2)}_{\mathcal{F},:} \quad \bm{0}_{(n-s)\times (n-d)}		\\
			E^{(r_s-1)}_{\bar{\mathcal{F}},:}  		\\
			\end{array}
		\right], 
	\end{eqnarray}
	where $\bar{\mathcal{F}}$ denotes the set of first $n-s-(L+r_s(n-d)-r_s(n-s))=n-s-L+r_s(d-s)$ elements in set $\mathcal{F}$.

  Denote $W = (E_{\mathcal{H},\mathcal{H}}^{(\mathcal{F})})^T\cdot E_{\mathcal{H},\mathcal{H}}^{(\mathcal{F})}$. Then, according to \cite{ChenD05}, the 2-norm condition number of matrix $E_{\mathcal{H},\mathcal{H}}^{(\mathcal{F})}$ can be written as
\begin{equation}\label{eqn:condition_num}
	\kappa  \triangleq \sqrt{\lambda_{max}\over \lambda_{min}},
\end{equation} 
where $\lambda_{max}, \lambda_{min}$ is the maximum and minimum eigenvalue of $W$, respectively. To simplify the notation, we denote $\hat{E} \triangleq E_{\mathcal{H},\mathcal{H}}^{(\mathcal{F})}$, $\hat{E}^{(i)} \triangleq E^{(i)}_{\mathcal{F},:}$ for $i\in[0:r_s-2]$, and $\hat{E}^{(r_s-1)} \triangleq E^{(r_s-1)}_{\bar{\mathcal{F}},:}$ for $i=r_s-1$. Then, by means of \eqref{eqn:E}, the matrix $W$ can also be rewritten as
\begin{equation}\label{eqn:W}
	W = \hat{E}^T\cdot \hat{E} = \sum_{i=0}^{r_s-1}
	\left[
		\begin{array}{ccccccccc}
			(\hat{E}^{(i)})^T\cdot \hat{E}^{(i)}  		&  \bm{0}_{(L+(i+1)(n-d)\times (r_s-i-1)(n-d)}	\\
			\bm{0}_{(r_s-i-1)(n-d)\times (L+(i+1)(n-d))}		& 	\bm{0}_{(r_s-i-1)(n-d)\times (r_s-i-1)(n-d))}		\\
			\end{array}
		\right].
\end{equation}

Denote $W^{(i)}=(\hat{E}^{(i)})^T\cdot \hat{E}^{(i)}$. Then, according to \cite{Edelman89}, $W^{(i)}$ is a Wishart matrix with parameters $(n-s,L+(i+1)(n-d))$ for $i\in[0:r_s-2]$ and a Wishart matrix with parameters $(n-s-L+r_s(d-s),L+r_s(n-d))$ for $i=r_s-1$.
Let $\lambda^{(i)}_{max}, \lambda^{(i)}_{min}$ be the maximum and minimum eigenvalue of $W^{(i)}$, respectively.

Then, by \eqref{eqn:W}, we have
\begin{eqnarray}\label{eqn:lambda_max}
	\lambda_{max} &=& \max_{||\textbf{u}||_2=1,\textbf{u}\in\mathbb{R}^{L+r_s(n-d)}} \textbf{u}^T\cdot W\cdot \textbf{u}	\nonumber \\ 
	&\leq & \sum_{i=0}^{r_s-1} \max_{||\textbf{u}^{(i)}||_2\leq1} (\textbf{u}^{(i)})^T\cdot W^{(i)}\cdot \textbf{u}^{(i)}	 \nonumber \\
	&= & \sum_{i=0}^{r_s-1}\lambda^{(i)}_{max},
\end{eqnarray}
where $\textbf{u}^{(i)}$ is the vector formed by the first $L+(i+1)(n-d)$ elements in $\textbf{u}$. Similarly,
\begin{eqnarray}\label{eqn:lambda_min}
	\lambda_{min} &=& \min_{||\textbf{u}||_2=1,\textbf{u}\in\mathbb{R}^{L+r_s(n-d)}} \textbf{u}^T\cdot W \cdot \textbf{u}	\nonumber \\
	&=&  \min_{||\textbf{u}^{(i)}||_2\leq1,i\in[0:r_s)} \sum_{i=0}^{r_s-1}(\textbf{u}^{(i)})^T\cdot W^{(i)}\cdot \textbf{u}^{(i)} 	\nonumber \\
	&\overset{(a)}{\geq} & \min_{||\textbf{u}^{(r_s-1)}||_2\leq 1} (\textbf{u}^{(r_s-1)})^T\cdot W^{(r_s-1)}\cdot \textbf{u}^{(r_s-1)}	\nonumber \\
	&= & \lambda^{(r_s-1)}_{min},
\end{eqnarray}
where $(a)$ follows from the fact that matrix $W^{(i)}$ is positive semidefinite such that its eigenvalues are non-negative,
i.e., $(\textbf{u}^{(i)})^T\cdot W^{(i)}\cdot \textbf{u}^{(i)}$ is non-negative for all $0\le i< r_s$.
Then, by \eqref{eqn:condition_num}, \eqref{eqn:lambda_max} and \eqref{eqn:lambda_min}, we have
\begin{equation*}
	\kappa \leq \sqrt{\sum_{i=0}^{r_s-1}\lambda^{(i)}_{max},\over \lambda_{min}^{(r_s-1)}}.
\end{equation*}

Therefore, given any real value $x$, the probability that the condition number of matrix $E_{\mathcal{H},\mathcal{H}}^{(\mathcal{F})}$ does not exceeds $x$ is 
\begin{eqnarray}\label{eqn:pr_k}
	\Pr(\kappa\leq x) &\geq & \Pr\left(\sqrt{\sum_{i=0}^{r_s-1}\lambda^{(i)}_{max},\over \lambda_{min}^{(r_s-1)}} \leq x\right) \nonumber \\
	&= & 1-	\Pr\left(\sqrt{\sum_{i=0}^{r_s-1}\lambda^{(i)}_{max},\over \lambda_{min}^{(r_s-1)}} > x\right)		\nonumber	\\
	&=& 1- \Pr\left(\log\left(\sum_{i=0}^{r_s-1}\lambda^{(i)}_{max},\over \lambda_{min}^{(r_s-1)}\right) > {2}\log x\right)	\nonumber \\
	&\overset{(a)}{\geq} & 1- {\mathbb{E}\left(\log\left(\sum_{i=0}^{r_s-1}\lambda^{(i)}_{max},\over \lambda_{min}^{(r_s-1)}\right)\right) \over {2}\log x}	\nonumber \\
	&=& 1- {\mathbb{E}\left(\log\left(\sum_{i=0}^{r_s-1}\lambda^{(i)}_{max}\right)\right) - \mathbb{E}(\log(\lambda_{min}^{(r_s-1)})) \over {2}\log x}	\nonumber \\
	& \overset{(b)}{\geq} & 1- {\log\left(\sum_{i=0}^{r_s-1}\mathbb{E}\left(\lambda^{(i)}_{max}\right)\right) - \mathbb{E}(\log(\lambda_{min}^{(r_s-1)})) \over {2}\log x},	
\end{eqnarray}
where $(a)$ follows from the Markov's inequality and $(b)$ follows from the Jensen Inequality. 

Notice that, given any $0<\epsilon<1$, by \eqref{eqn:pr_k}, if 
\begin{equation}\label{eqn:logx_lb}
	2\epsilon \log{x}\geq  \log\left(\sum_{i=0}^{r_s-1}\mathbb{E}\left(\lambda^{(i)}_{max}\right)\right) - \mathbb{E}(\log(\lambda_{min}^{(r_s-1)})),
\end{equation}
then $\Pr(\kappa \leq x)\geq 1-\epsilon$. That is, the probability that the condition number being smaller than $x$ is at least $1-\epsilon$ if \eqref{eqn:logx_lb} is satisfied.

To attain \eqref{eqn:logx_lb}, we first give two useful lemmas from the literature.

\begin{Lemma}[Based on the Proposition 5.1 in \cite{Edelman89}]
	Given a random matrix $E$ with size of $u_r\times u_c$, whose entries are i.i.d. standard Gaussian, if $W=EE^T$ follows the Wishart distribution with parameter $(u_r,u_c)$ and $\lambda_{min}$ is the minimum eigenvalue of $W$, where $\lim_{u_c\rightarrow \infty}{u_r \over u_c}=y,0<y<1$, then, 
	\begin{equation*}
		\mathbb{E}(\log{\lambda_{min}}) = \log{u_c}+\log{(1-\sqrt{y})^2}+O(1).
	\end{equation*}
\end{Lemma}

\begin{Lemma}[Based on the equations (32) and (42) in \cite{Chiani14}]
	Given random matrix $E$ with size of $u_r\times u_c$, whose entries are i.i.d. standard Gaussian, and $\lambda_
	{max}$ is the maximum eigenvalue of $EE^T$, then, for $u_r,u_c \rightarrow \infty$ and ${u_r\over u_c}\rightarrow y\in[0,\infty]$,
	\begin{equation*}
		{\lambda_{max} - \mu_{u_r,u_c} \over \sigma_{u_r,u_c}} \overset{\mathcal{D}}{\rightarrow} \mathcal{TW} \simeq \mathcal{G}-\alpha,
	\end{equation*}
	where $\mathcal{TW}$ denotes a random variable (r.v.) with Tracy-Widom distribution of order $1$, 
	\begin{equation}\label{eqn:u_sigma}
		\mu_{u_r,u_c}=(\sqrt{u_r-1/2}+\sqrt{u_c-1/2})^2, \quad \sigma_{u_r,u_c} = \sqrt{\mu_{u_r,u_c}}\left({1\over \sqrt{u_r-1/2}}+{1\over \sqrt{u_c-1/2}}\right)^{1\over 3}=\mu_{u_r,u_c}^{1\over 3},
	\end{equation}
$\alpha \approx 9.848$, $\mathcal{G}\sim \Gamma(\gamma,\theta)$ denotes a Gamma r.v. with shape parameter $\gamma\approx 46.446$, scale parameter $\theta \approx 0.186054$, and $\mathbb{E}(\mathcal{G})=\gamma\theta \approx 8.6415$.
\end{Lemma}

Note that the parameters $u_r,u_c$ in our problem are 
 \begin{equation*}
 	u_r^{(i)} = n-s, \quad u_c^{(i)} = L+(i+1)(n-d),
 \end{equation*} 
 for $i\in[0:r_s-2]$ and
  \begin{equation*}
 	u_r^{(r_s-1)} = n-s-L+r_s(d-s), \quad u_c^{(r-1)} = L+r_s(n-d).
 \end{equation*} 
Then, based on Lemmas 1 and 2, when $n$ is large enough, we have $u_r^{(i)},u_c^{(i)}\rightarrow \infty$. Therefore, the condition in \eqref{eqn:logx_lb} can be further simplified as:
\begin{eqnarray}\label{eqn:x_simplified}
x &\geq & e^{{1\over 2\epsilon}\left(\log{\sum_{i=0}^{r_s-1}\mathbb{E}\left(\lambda^{(i)}_{max}\right)} -\mathbb{E}(\log{\lambda_{min}^{(r_s-1)}}) \right) }	\nonumber \\
& \overset{(a)}{\approx} & e^{{1\over 2\epsilon} \left(\log{\sum_{i=0}^{r_s-1}\mathbb{E}\left(\lambda^{(i)}_{max}\right)} - \log{u_c^{(r_s-1)}}-\log{(\sqrt{r_s}-1)^2\over r_s}-O(1) \right)}	\nonumber\\
& \overset{(b)}{=} & \left(r_s\sum_{i=0}^{r_s-1}\mathbb{E}\left(\lambda^{(i)}_{max}\right) \over u_c^{(r_s-1)}(\sqrt{r_s}-1)^2\cdot \beta \right)^{1\over 2\epsilon}	\nonumber\\
& \overset{(c)}{=} & \left( r_s\sum_{i=0}^{r_s-1}\sigma_{u_r^{(i)},u_c^{(i)}}\left(\mathbb{E}\left({\lambda^{(i)}_{max}-\mu_{u_r^{(i)},u_c^{(i)}} \over \sigma_{u_r^{(i)},u_c^{(i)}}}\right) + \mu_{u_r^{(i)},u_c^{(i)}}\right) \over u_c^{(r_s-1)}(\sqrt{r_s}-1)^2\cdot \beta \right)^{1\over 2\epsilon} \nonumber	\\
& \overset{(d)}{\approx} & \left( r_s\sum_{i=0}^{r_s-1}\sigma_{u_r^{(i)},u_c^{(i)}}\left(\mathbb{E}\left(\mathcal{G}-\alpha \right) + \mu_{u_r^{(i)},u_c^{(i)}}\right) \over u_c^{(r_s-1)}(\sqrt{r_s}-1)^2\cdot \beta \right)^{1\over 2\epsilon}. \nonumber	\\
& \overset{(e)}{=} & \left( r_s\sum_{i=0}^{r_s-1}\sigma_{u_r^{(i)},u_c^{(i)}}\left( \mu_{u_r^{(i)},u_c^{(i)}} - 1.2065 \right) \over u_c^{(r_s-1)}(\sqrt{r_s}-1)^2\cdot \beta \right)^{1\over 2\epsilon}, 
\end{eqnarray}
where $(a)$ is based on Lemma 1 and $y=\lim_{u_c\rightarrow\infty}{u_r^{(r_s-1)}\over u_c^{(r_s-1)}}=\lim_{n\rightarrow\infty}{n-s+L-r_s(d-s)\over L+r_s(n-d)}={1\over r_s}$, $\beta\triangleq e^{O(1)}$ is a constant in $(b)$, $(c)$ and $(d)$ are based on Lemma 2, and $(e)$ is based on the fact $\mathbb{E}(\mathcal{G}-\alpha)=\mathbb{E}(\mathcal{G})-\alpha$. 

That is, when $x$ is larger than \eqref{eqn:x_simplified}, for some $\epsilon\in(0:1)$, we can have the probability that the condition number of matrix $E_{\mathcal{H},\mathcal{H}}^{(\mathcal{F})}$ not exceeding $x$ is at least $1-\epsilon$, i.e.,
\begin{equation}\label{eqn:k_prob}
	\Pr(\kappa\leq x) \geq 1-\epsilon.
\end{equation}

To see the scaling law with $n$, we derive an upper bound for the right-hand side of \eqref{eqn:x_simplified} as
\begin{eqnarray}\label{eqn:x_lb}
\left( r_s\sum_{i=0}^{r_s-1}\sigma_{u_r^{(i)},u_c^{(i)}}\left( \mu_{u_r^{(i)},u_c^{(i)}} - 1.2065 \right) \over u_c^{(r_s-1)}(\sqrt{r_s}-1)^2\cdot \beta \right)^{1\over 2\epsilon} 	
&\overset{(a)}{<}& \left( r_s\sum_{i=0}^{r_s-1} \mu_{u_r^{(r_s-1)},u_c^{(i)}}^{4\over 3} \over u_c^{(r_s-1)}(\sqrt{r_s}-1)^2\cdot \beta \right)^{1\over 2\epsilon} 	\nonumber \\
&\overset{(b)}{<}& \left( r_s^2\cdot (4u_c^{(r_s-1)})^{4\over 3} \over u_c^{(r_s-1)}(\sqrt{r_s}-1)^2\cdot \beta\right)^{1\over 2\epsilon} 	\nonumber \\
&=& \left( 4^{4\over 3} r_s^2 \over (\sqrt{r_s}-1)^2\cdot \beta\right)^{1\over 2\epsilon}\cdot (u_c^{(r_s-1)})^{1\over 6\epsilon}\nonumber \\
&=& \left( 4^{4\over 3} r_s^2 \over (\sqrt{r_s}-1)^2\cdot \beta\right)^{1\over 2\epsilon}\cdot (L+r_s(n-d))^{1\over 6\epsilon},
\end{eqnarray} 
where $(a)$ is due to the fact that $\sigma_{u_r^{(i)},u_c^{(i)}}(\mu_{u_r^{(i)},u_c^{(i)}} - 1.2065) < \sigma_{u_r^{(i)},u_c^{(i)}}\mu_{u_r^{(i)},u_c^{(i)}} $ and \eqref{eqn:u_sigma}, and $(b)$ is due to the fact that  $\mu_{u_r^{(i)},u_c^{(i)}}<\left(2\sqrt{u_c^{(i)}-1/2}\right)^2<\left(2\sqrt{u_c^{(i)}}\right)^2=4u_c^{(i)}\leq 4u_c^{(r-1)}$ for $i\in[0:r_s-1)$ because
$u_c^{(i)}$ is a monotonically increasing in $i$ and $u_c^{(i)}\geq u_r^{(i)}$.
Then, based on \eqref{eqn:x_simplified}, when $x$ is set to be the R.H.S. of \eqref{eqn:x_lb}, \eqref{eqn:k_prob} is also satisfied. 

That is, given fixed constants $L,d$ and any $s\in[0:d-1],r_s=\lceil{L\over d-s}\rceil$, when $n$ is large enough, the condition number of matrix $E_{\mathcal{H},\mathcal{H}}^{(\mathcal{F})}$ is in order $O(n^{1\over 6\epsilon})$ for any $0<\epsilon<1$. Particularly, when $\epsilon=1/6$, there is at least $1-\epsilon\approx 83\%$ probability that the condition of matrix $E_{\mathcal{H},\mathcal{H}}^{(\mathcal{F})}$ is bounded by a linear function of $n$.


\textbf{\textit{Proof of \eqref{eqn:numerical_agc}}:}
\begin{enumerate}
\item If $i=0$, for each $s\in[0:s_{\max}]$,
\begin{IEEEeqnarray}{rCl}
&& h(u,0,s,s_{\max}) \notag \\
&=&\Pr(Y=0,S_0=s) \notag \\
&=&\Pr(S_0=s)\Pr(Y=0\,|\,S_0=s) \notag \\
&=&\Pr(S_0=s)\label{eq:1p}\\
&=&{u\choose s}p^s(1-p)^{u-s}, \notag
\end{IEEEeqnarray}
where \eqref{eq:1p} follows from the fact $\Pr(Y=0\,|\,S_0=0)=1$.
\item If $i\geq 1$, then for each $s\in[0:s_{\max}]$,
\begin{IEEEeqnarray}{rl}
&\quad h(u,i,s,s_{\max})\notag \\
&=\Pr(Y=i,S_i=s) \notag \\
&=\Pr(S_0\geq S_1\geq\ldots\geq S_{i-1}>s_{\max},S_i=s) \notag \\
&=\sum_{s_{\max}<a_{i\!-\!1}\leq \ldots\leq a_0\leq u}\Pr(S_0\!=\!a_0,\ldots, S_{i\!-\!1}\!=\!a_{i-1},S_i\!=\!s) \notag	\\ 
&=\sum_{s_{\max}<a_{i\!-\!1}\leq \ldots\leq a_0\leq u}\Pr(S_0=a_0)	\notag\\
&~ \! \! \!\! \cdot \prod_{j=1}^{i\!-\!1} \Pr( S_j\!=\!a_j|S_{j\!-\!1}\!=\!a_{j\!-\!1}) \! \Pr(S_{i}\!=\!s|S_{i\!-\!1}\!=\!a_{i\!-\!1})\label{eqn:numerical:S:markov}
\\
&=\sum_{s_{\max}<a_{i-1}\leq \ldots\leq a_0\leq u}{u\choose a_0}p^{a_0}(1-p)^{u-a_0}	\notag\\
& \quad \!\! \!\cdot \prod_{j=1}^{i-1}{a_{j-1}\choose a_j}p^{a_j}(1-p)^{a_{j-1}\!-\!a_j}{a_{i\!-\!1}\choose s}p^{s}(1-p)^{a_{i\!-\!1}-s} \notag \\
 &=\sum_{s_{\max}<a_{i\!-\!1}\leq \ldots\leq a_0\leq u}\frac{u!}{(u\!-\!a_0)!(a_0\!-\!a_1)!\ldots(a_{i-1}\!-\!s)! s!} \notag\\
 &\quad \cdot p^{\sum_{j=0}^{i-1}a_j+s}(1-p)^{u-s}, \notag
\end{IEEEeqnarray}

where \eqref{eqn:numerical:S:markov} holds because of \eqref{Eqn_Markov}, i.e., $S_0,S_1,\ldots,$ forms a Markov chain.
\end{enumerate}


\textbf{\textit{Proof of \eqref{eqn:eva}}:}

Firstly, we can have:
\begin{eqnarray}
&&\Pr(Y^{\textnormal{G-A}}=i,S^{\textnormal{G-A}}_i=s)\notag\\
&=&\Pr(Y^{\textnormal{G-A}}\leq i,S^{\textnormal{G-A}}_i\leq s)\!-\!\Pr(Y^{\textnormal{G-A}}\leq i,S^{\textnormal{G-A}}_i\leq s\!-\!1) -\Pr(Y^{\textnormal{G-A}}\leq i-1,S^{\textnormal{G-A}}_i\leq s)+\Pr(Y^{\textnormal{G-A}}\leq i-1,S^{\textnormal{G-A}}_i\leq s\!-\!1)\notag\\
&=&\hspace{-3mm}\Pr(Y^{\textnormal{G-A}}\leq i,S^{\textnormal{G-A}}_i\leq s)\!-\!\Pr(Y^{\textnormal{G-A}}\leq i,S^{\textnormal{G-A}}_i\leq s\!-\!1) -\Pr(Y^{\textnormal{G-A}}\leq i-1)+\Pr(Y^{\textnormal{G-A}}\leq i-1)\! \cdot\! \mathbf{1}(s\geq 1)\label{step:drop}\\
&=&\hspace{-3mm}\Pr(Y^{\textnormal{G-A}}\leq i,S^{\textnormal{G-A}}_i\leq s)\!-\!\Pr(Y^{\textnormal{G-A}}\leq i,S^{\textnormal{G-A}}_i\leq s\!-\!1)-\Pr(Y^{\textnormal{G-A}}\leq i-1)\cdot \mathbf{1}(s=0)\notag\\
&=&\hspace{-3mm}\left\{\begin{array}{ll}
\hspace{-2mm}\Pr(Y^{\textnormal{G-A}}\!\leq \!i,S^{\textnormal{G-A}}_i\!=\!0)\!-\\
\quad \!\Pr(Y^{\textnormal{G-A}}\!\leq\! i\!-\!1),&\textnormal{if}~s=0\\
\hspace{-2mm}\Pr(Y^{\textnormal{G-A}}\!\leq \!i,S^{\textnormal{G-A}}_i\!\leq\! s)\!-\\
\quad \!\Pr(Y^{\textnormal{G-A}}\!\leq\! i,S^{\textnormal{G-A}}_i\!\leq\! s-1),&\textnormal{if}~1\!\leq\! s\!\leq\! s_{\max}^{\textnormal{A}}
\end{array}
\right.\notag
\end{eqnarray}
where \eqref{step:drop} follows that given $Y^{\textnormal{G-A}}\leq i-1$, it must hold $S_i^{\textnormal{G-A}}=0$ since no worker will be restarted in epoch $i$.
Therefore, to obtain \eqref{eqn:eva}, we need to evaluate the probabilities $\Pr(Y^{\textnormal{G-A}}\leq i,S^{\textnormal{G-A}}_i=0),\Pr(Y^{\textnormal{G-A}}\leq i-1) $ for the case $s=0$, and $\Pr(Y^{\textnormal{G-A}}\leq i,S^{\textnormal{G-A}}_i\leq s),\Pr(Y^{\textnormal{G-A}}\leq i,S^{\textnormal{G-A}}_i\leq s-1)$ for the case $1\leq s\leq s_{\max}^{\rm A}$.

For the case $s=0$,
\begin{IEEEeqnarray}{rl}
&\quad \Pr(Y^{\textnormal{G-A}}\leq i,S^{\textnormal{G-A}}_i=0) \notag \\
&=\Pr\Big(\max_{j\in[0:m]}\{Y_j^{\textnormal{G-A}}\}\leq i,\max_{j\in[0:m]}\{S^{\textnormal{G-A}}_{j,i}\}=0\Big) \notag \\
&=\Pr\big(Y_j^{\textnormal{G-A}}\leq i,S^{\textnormal{G-A}}_{j,i}=0, {\forall}\,j\in[0:m]\big) \notag \\
&=\prod_{j=0}^m \Pr\big (Y_j^{\textnormal{G-A}}\leq i,S^{\textnormal{G-A}}_{j,i}=0\big)\label{eqn:ind:a}\\
&=\prod_{j=0}^m \Big(\sum_{b=0}^i\Pr\big (Y_j^{\textnormal{G-A}}= b,S^{\textnormal{G-A}}_{j,i}=0\big)\Big) \notag \\
&=\prod_{j=0}^m \Big(\sum_{b=0}^{i-1}\Pr\big (Y_j^{\textnormal{G-A}}= b\big)+\Pr\big (Y_j^{\textnormal{G-A}}= i,S^{\textnormal{G-A}}_{j,i}=0\big)\Big)\IEEEeqnarraynumspace\label{step:drop:S}\\
&=\prod_{j=0}^m \Big(\sum_{b=0}^{i-1}\sum_{a=0}^{s_{\max}^{\textnormal{A}}}\Pr\big (Y_j^{\textnormal{G-A}}= b,S_{j,b}^{\textnormal{G-A}}=a\big) +\Pr\big (Y_j^{\textnormal{G-A}}= i,S^{\textnormal{G-A}}_{j,i}=0\big)\Big) \notag \\
&=\Big(\sum_{b=0}^{i-1}\sum_{a=0}^{s_{\max}^{\textnormal{A}}} h(d,b,a,s_{\max}^{\textnormal{A}})+h(d,i,0,s_{\max}^{\textnormal{A}})\Big)^m \cdot \Big(\sum_{b=0}^{i-1}\sum_{a=0}^{s_{\max}^{\textnormal{A}}} h(n\!-\!md,b,a,s_{\max}^{\textnormal{A}})\!+\!h(n\!-\!md,i,0,s_{\max}^{\textnormal{A}})\!\Big),\label{step:plug:h:a}
\end{IEEEeqnarray}
where \eqref{eqn:ind:a} holds since different groups of workers work independently; \eqref{step:drop:S} is true due to the fact $\Pr(S_{j,i}^{\textnormal{G-A}}=0\,|\,Y_j^{\textnormal{G-A}}=b<i)=1$; and \eqref{step:plug:h:a} follows from the definition of the function $h$. 
And the second probability
\begin{IEEEeqnarray}{rCl}
&&\Pr(Y^{\textnormal{G-A}}\leq i-1) \notag \\
&=&\Pr\Big(\max_{j\in[0:m]}\{Y_j^{\textnormal{G-A}}\}\leq i-1\Big) \notag \\
&=&\Pr\big(Y_j^{\textnormal{G-A}}\leq i-1\,,\forall\,j\in[0:m]\big) \notag \\
&=&\prod_{j=0}^m \Pr\big (Y_j^{\textnormal{G-A}}\leq i-1\big)\label{eqn:ind:b}\\
&=&\prod_{j=0}^m \Big(\sum_{b=0}^{i-1}\Pr\big (Y_j^{\textnormal{G-A}}= b\big)\Big) \notag \\
&=&\prod_{j=0}^m \Big(\sum_{b=0}^{i-1}\sum_{a=0}^{s_{\max}^{\textnormal{A}}}\Pr\big (Y_j^{\textnormal{G-A}}= b,S_{j,b}^{\textnormal{G-A}}=a\big)\Big) \notag \\
&=&\Big(\sum_{b=0}^{i-1}\sum_{a=0}^{s_{\max}^{\textnormal{A}}} h(d,b,a,s_{\max}^{\textnormal{A}})\Big)^m  \cdot \Big(\sum_{b=0}^{i-1}\sum_{a=0}^{s_{\max}^{\textnormal{A}}} h(n-md,b,a,s_{\max}^{\textnormal{A}})\Big),\label{step:plug:h:b}
\end{IEEEeqnarray}
where \eqref{eqn:ind:b} holds since different groups of workers work independently;  and \eqref{step:plug:h:b} follows from the definition of the function $h$.

 For the case $1\leq s\leq s_{\max}^{\textnormal{A}}$\footnote{In the evaluations, we adopt the convention that $\sum_{x=a}^bf(x)=0$ if $b<a$ for any function $f(x)$.},
\begin{IEEEeqnarray}{rCl}
&&\Pr(Y^{\textnormal{G-A}}\leq i,S^{\textnormal{G-A}}_i\leq s) \notag \\
&=&\Pr\Big(\max_{j\in[0:m]}\{Y_j^{\textnormal{G-A}}\}\leq i,\max_{j\in[0:m]}\{S_{j,i}^{\textnormal{G-A}}\}\leq s\}\Big)\label{eq:first}\\
&=&\Pr\big(Y_j^{\textnormal{G-A}}\leq i,S_{j,i}^{\textnormal{G-A}}\leq s,\forall\, j\in[0:m]\big) \notag \\
&=&\prod_{j=0}^m\Pr(Y_j^{\textnormal{G-A}}\leq i,S_{j,i}^{\textnormal{G-A}}\leq s)\label{eqn:ind}\\
&=&\prod_{j=0}^m\bigg(\sum_{b=0}^{i-1}\Pr(Y_j^{\textnormal{G-A}}=b,S_{j,i}^{\textnormal{G-A}}\leq s) +\sum_{a=0}^s\Pr(Y_j^{\textnormal{G-A}}=i,S_{j,i}^{\textnormal{G-A}}=a)\bigg) \notag \\
&=&\prod_{j=0}^m\bigg(\sum_{b=0}^{i-1}\Pr(Y_j^{\textnormal{G-A}}=b) +\sum_{a=0}^s\Pr(Y_j^{\textnormal{G-A}}=i,S_{j,i}^{\textnormal{G-A}}=a)\bigg)\label{eq:drop:Sji}\\
&=&\prod_{j=0}^m\bigg(\sum_{b=0}^{i-1}\sum_{a=0}^{s_{\max}^{\textnormal{A}}}\Pr(Y_j^{\textnormal{G-A}}=b,S_{j,b}^{\textnormal{G-A}}=a) +\sum_{a=0}^s\Pr(Y_j^{\textnormal{G-A}}=i,S_{j,i}^{\textnormal{G-A}}=a)\bigg) \notag \\
&=&\bigg(\sum_{b=0}^{i-1}\sum_{a=0}^{s_{\max}^{\textnormal{A}}}h(d,b,a,s_{\max}^{\textnormal{A}})+\sum_{a=0}^{s}h(d,i,a,s_{\max}^{\textnormal{A}})\bigg)^m \cdot \bigg(\!\sum_{b=0}^{i-1}\sum_{a=0}^{s_{\max}^{\textnormal{A}}}\!h(n\!-\!md,b,a,s_{\max}^{\textnormal{A}}) +\sum_{a=0}^{s}h(n\!-\!md,i,a,s_{\max}^{\textnormal{A}})\!\bigg),\label{eq:plug:h}
\end{IEEEeqnarray}
where \eqref{eqn:ind} holds since different groups of workers work independently; \eqref{eq:drop:Sji} is  true due to the fact $\Pr(S_{j,i}^{\textnormal{G-A}}\leq s\,|\,Y_j^{\textnormal{G-A}}=b<i)=1$; and \eqref{eq:plug:h} follows from the definition of the function $h$. 
Following the same steps as \eqref{eq:first}--\eqref{eq:plug:h}, 
\begin{IEEEeqnarray*}{rCl}
&&\Pr(Y^{\textnormal{G-A}}\leq i,S^{\textnormal{G-A}}_i\leq s-1)\\
&=&\bigg(\sum_{b=0}^{i-1}\sum_{a=0}^{s_{\max}^{\textnormal{A}}}h(d,b,a,s_{\max}^{\textnormal{A}})+\sum_{a=0}^{s-1}h(d,i,a,s_{\max}^{\textnormal{A}})\bigg)^m \cdot \bigg(\sum_{b=0}^{i-1}\sum_{a=0}^{s_{\max}^{\textnormal{A}}}h(n-md,b,a,s_{\max}^{\textnormal{A}}) +\sum_{a=0}^{s-1}h(n-md,i,a,s_{\max}^{\textnormal{A}})\bigg).
\end{IEEEeqnarray*}


\textbf{\textit{Proof of Theorem \ref{thm:group_lb}:}}

Suppose that the dataset is divided into $k$ data subsets and each worker computes $\gamma$ data subsets, the workers are partitioned into $h$ groups. Denote $n_i$ and $k_i$ the number of workers and data subsets in group $i\in[0:h)$, respectively. Then, we have
\begin{equation*}
	\sum_{i=0}^{h-1}k_i=k, \quad \sum_{i=0}^{h-1}n_i = n.
\end{equation*}
Since each group computes distinct data subsets, and each data subset is computed $d$ times by different workers, it is easy to know $n_i\geq d$ and 
\begin{equation*}
	\gamma n_i = k_i d.
\end{equation*}
Then, for each group $i\in[0:h)$, it needs at least $\lceil {k_i \over \gamma} \rceil = \lceil {n_i\over d} \rceil$ active workers to compute the full $k_i$ data subsets. Therefore, $s_{\rm{total}}^{\ast}$ is upper bounded by
\begin{eqnarray*}
	s_{\rm{total}}^{\ast} &\leq& \sum_{i=0}^{h-1}\left(n_i- \Big\lceil {n_i\over d}  \Big\rceil\right) \\
	&=& \sum_{i=0}^{h-1}\left(n_i - {n_i+\delta_i \over d}\right)	\\
	&=& n-{n+\sum_{i=0}^{h-1}\delta_i\over d} \\
	&=& n-{n\over d} - {\sum_{i=0}^{h-1}\delta_i \over d},
\end{eqnarray*}
where $\delta_i= \lceil {n_i\over d} \rceil d- n_i \in[0:d)$, and $\sum_{i=0}^{h-1}\delta_i$ is lower bounded by 
\begin{equation*}
	\sum_{i=0}^{h-1}\delta_i=\sum_{i=0}^{h-1} \Big\lceil{n_i\over d} \Big\rceil d-n \geq  \Big\lceil {\sum_{i=0}^{h-1}n_i \over d} \Big\rceil d - n  = \Big\lceil {n\over d}  \Big\rceil d-n.
\end{equation*}
Then,
\begin{equation}\label{eqn:group:lb}
	s_{\rm{total}}^{\ast} \leq n-{n\over d}-{\sum_{i=0}^{h-1}\delta_i \over d} \leq n- \Big\lceil{n\over d}\Big\rceil.
\end{equation}

Consider the G-AGC scheme, which partitions the nodes into $\lfloor\frac{n}{d}\rfloor$ groups according to \eqref{eqn:Ij}, where each group can tolerate $d-1$ stragglers. Thus, 
\begin{equation}
s_{\rm{total}}^{\rm{G\text{-}AGC}}=\Big\lfloor\frac{n}{d}\Big\rfloor(d-1). \notag
\end{equation}
Let $n=\theta\cdot d+\lambda$, where $\theta=\lfloor \frac{n}{d}\rfloor$, and $0\leq \lambda<d$, then
\begin{eqnarray}
	s_{\rm{total}}^{\ast} - s_{\rm{total}}^{\rm{G\text{-}AGC}} &\leq&  n - \Big\lceil {n\over d} \Big\rceil - \Big\lfloor {n\over d}\Big\rfloor (d-1) \notag\\
&=&\theta d+\lambda-\Big\lceil \frac{\theta d+\lambda}{d}\Big\rceil-\theta(d-1)\notag\\
&=&\lambda-\Big\lceil\frac{\lambda}{d}\Big\rceil\label{eq:lambda}\\
&<&d-1\notag.
\end{eqnarray} 
This proves the bound $s_{\rm{total}}^{\rm{G\text{-}AGC}}>s_{\rm{total}}^{\ast}-d+1$. Moreover, by \eqref{eq:lambda}, if $\lambda=0$ or $1$, $ s_{\rm{total}}^{\rm{G\text{-}AGC}}= s_{\rm{total}}^{\ast} $, i.e., when $d|n$ or $n=\theta d+1$ for some $\theta\geq 1$,  G-AGC scheme achieves the optimal total number of stragglers across the groups among all group schemes satisfying C1--C3.

\end{appendix}

\end{document}